\documentclass{article}
\usepackage[hyphens]{url}
\usepackage[utf8]{inputenc}
\usepackage{graphicx}
\usepackage{authblk}
\usepackage[english]{babel}
\usepackage{setspace}
\usepackage{fancyhdr}
\usepackage{array}
\usepackage[margin=1in]{geometry}
\usepackage{float}
\usepackage{amsmath}
\usepackage{amssymb}
\usepackage{bbm}
\usepackage{bm}
\usepackage{scalerel,stackengine}
\usepackage[final]{pdfpages}
\usepackage{longtable}
\usepackage{chngcntr}
\usepackage{placeins}
\usepackage{microtype}
\usepackage{tikz}
\usepackage{makecell}
\usepackage{hyperref}
\usepackage{subcaption}
\usepackage{rotating}
\usepackage{soul}
\usepackage{tabularx} 
\usepackage{pdfpages}
\usepackage[square,numbers]{natbib}
\usepackage{multirow}
\usepackage{pdflscape}

\hypersetup{
    colorlinks = true,
    citecolor = {blue},
    urlcolor = {blue},
    menucolor = {blue},
    linkcolor = {blue}
}

\makeatletter
\patchcmd{\@maketitle}{\LARGE \@title}{\fontsize{18}{20}\selectfont\textbf{\@title}}{}{}
\makeatother

\title{Statistical modeling to adjust for time trends in adaptive platform trials utilizing non-concurrent controls}

\author[1]{Pavla Krotka}
\author[1]{Martin Posch}
\author[2]{Mohamed Gewily}
\author[3,4,5,6]{Günter Höglinger}
\author[1]{Marta Bofill Roig\thanks{marta.bofillroig@meduniwien.ac.at}}
\affil[1]{Center for Medical Data Science, Medical University of Vienna, Vienna 1090, Austria}
\affil[2]{Department of Pharmacy, Uppsala University, Uppsala, Sweden.}
\affil[3]{Department of Neurology, LMU University Hospital, Ludwig-Maximilians-Universität (LMU) München, Munich, Germany.}
\affil[4]{German Center for Neurodegenerative Diseases (DZNE), Munich, Germany.}
\affil[5]{Munich Cluster for Systems Neurology (SyNergy), Munich, Germany.}
\affil[6]{Department of Neurology, Hannover Medical School, Hanover, Germany.}
\date{}         
\begin{document}
	
	\maketitle

\begin{abstract}
Utilizing non-concurrent control data (NCC) in the analysis of late-entering arms in platform trials has recently received considerable attention. While incorporating NCC can lead to increased power and lower sample sizes, it might introduce bias to the effect estimators if temporal drifts are present. Aiming to mitigate this potential bias, we propose various frequentist model-based approaches that leverage the NCC, while adjusting for time. One of the currently available models incorporates time as a categorical fixed effect, separating the trial duration into periods, defined as time intervals bounded by any arm entering or leaving the platform. In this work, we propose two extensions of this model. First, we consider an alternative definition of time by dividing the trial into fixed-length calendar time intervals. Second, we propose alternative model-based time adjustments. Specifically, we investigate adjusting for random effects and employing splines to model time with a polynomial function. We evaluate the performance of the proposed approaches in a simulation study and illustrate their use through a case study. We show that adjusting for time via a spline function controls the type I error in trials with a sufficiently smooth time trend pattern and may lead to power gains compared to the standard fixed effect model. However, the fixed effect model with period adjustment is the most robust model for arbitrary time trends, provided that the trend is equal across all arms. Especially, in trials with sudden changes in the time trend, the period-adjustment model is preferred if NCC are included.
\end{abstract}

\section{Introduction}\label{sect_intro}

Platform trials accelerate drug development by offering increased flexibility and efficiency \citep{Koenig2024Current}. They evaluate the efficacy of multiple treatment arms under a single master protocol, with the added benefit of permitting treatment arms to enter the trial over time and to stop early based on interim data \citep{Woodcock2017Master}. This design enables faster evaluation of drugs that are being developed, as they can be directly included to the shared platform \citep{Redman2015Master}. Moreover, efficiency gains can be achieved by sharing the control group across the whole platform. The sharing of resources lowers the number of patients required in the control arm as well as the overall number of patients needed for the trial compared to stand-alone trials. This poses an advantage also from the ethical and patient perspective \citep{Saville2016Efficiencies}. Given their complexity, designing platform trials requires extensive computer simulations and the use of software to assess their operating characteristics has become crucial for evaluating the robustness of the proposed design across different scenarios \citep{Meyer2020Evolution}. Moreover, valid statistical inference in platform trials remains of major concern to regulatory authorities, as several challenges arise due to the wide range of adaptive features, such as the late addition of experimental arms \citep{Lee2021Statistical}. 

\begin{figure}[!h]
    \centering
    \includegraphics[width=0.6\textwidth]{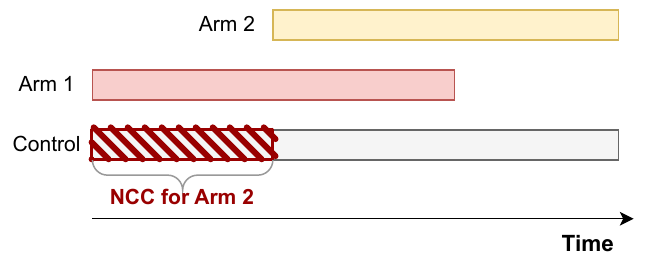}
    \caption{Non-concurrent control data for arm 2.}
    \label{fig:trial_scheme_NCC}
\end{figure}

In particular, the use of the common control group in the statistical analysis of added arms has been a subject of lively discussions \citep{Dodd2021Platform, Lee2020Including, Bofill2022Model, Saville2022Bayesian}. In platform trials, for treatment arms that enter when the trial is ongoing, the control group is divided into two separate groups: the concurrent controls (CC), which include patients that were randomized to the control arm at the same time as the given treatment arm was part of the platform trial and thus had a positive allocation probability of being randomized to the investigated treatment arm; and the non-concurrent controls (NCC), which denote patients randomized to the control group before the evaluated treatment arm entered the platform. Figure \ref{fig:trial_scheme_NCC} exemplifies a platform trial with two experimental arms, where the second arm joined the trial at a later time point. NCC data for the second arm are marked with a diagonal striped pattern. 
Including NCC data in treatment-control comparisons can offer several benefits, such as a lower required sample size or increased statistical power compared to analysis based on concurrent controls only. However, since treatment arms are added sequentially, randomization occurs at different times. This lack of true randomization over time might lead to an inflation of the type I error rate and can introduce bias in the treatment effect estimators due to time trends. Time trends can be caused, for example, by changes in the standard of care, patient population, or seasonal effects \citep{FDA2023Master}.

Recently, several modeling approaches were proposed to overcome the challenge of leveraging the NCC data, while still controlling the type I error rate and without introducing bias to the effect estimators. Focusing on a platform trial with two experimental arms, \cite{Lee2020Including} and \cite{Bofill2022Model} considered frequentist modeling approaches, where time is included to the model either as a linear function, or a categorical covariate. For the latter, the trial is split into two periods, separated by the time point where the second experimental arm was added to the platform. It was shown that both models improve the statistical power as compared to analysis using CC data only and control the type I error rate, if the functional form of the time trend is correctly specified, the time trends are equal across all arms and additive on the model scale \citep{Bofill2022Model}. Moreover, the model with categorical adjustment asymptotically maintains the type I error rate even under misspecification of the functional form of the time trend, provided that block randomization is used. To deal with situations where time trends could differ between arms, \cite{Bofill2022Model} also investigated a model that includes interaction between treatment and time. However, while this model maintains the type I error rate also in cases with different time trends, the power gain from using NCC data is lost.
A Bayesian alternative was suggested by \cite{Saville2022Bayesian} in their so-called Time Machine approach, which uses a Bayesian hierarchical model and adjusts for time in terms of intervals (``buckets") of equal length. The model smooths over the control response across the time buckets, such that closer buckets are modeled with more similar responses. They show that the Time Machine approximately controls the type I error rate and leads to superior performance in terms of statistical power as compared to the frequentist model with categorical adjustment for time in scenarios with linear time trend pattern. 
\cite{Marschner2022Analysis} proposed to analyze platform trials using meta-analysis techniques, which allow to conduct both treatment-control and treatment-treatment comparisons for arms that are not simultaneously active in the trial. In this network meta-analysis approach, the platform trial is viewed as a network of direct (concurrent) comparisons, made within periods between trial adaptations; and indirect (non-concurrent) comparisons, made across multiple stages in the trial using a common reference arm. The estimates of the non-concurrent treatment-control comparisons are then obtained by linearly combining the direct contrast estimates. 

In this work, we extend the frequentist modeling strategies from \cite{Bofill2022Model}. In particular, we propose an alternative definition of the time variable and consider adjusting for equally spaced calendar time intervals, similarly to the Bayesian Time Machine, also in the frequentist setting. Moreover, we investigate more advanced modeling approaches, such as mixed models or polynomial splines. The assumption of equal time trends, required for all methods proposed so far, may not always hold in practice, as temporal changes may affect treatment arms differently, e.g. in settings with different disease variants. Therefore, we investigate whether this assumption can be relaxed by considering a mixed model, which includes the interaction between treatment and time as a random effect. The performance of the proposed methods is evaluated in a simulation study, where we also assess the conditions under which the approaches lead to valid statistical inference. The use of the considered methods is demonstrated on a real-world dataset from a clinical trial for a rare disease (progressive supranuclear palsy), and in a hypothetical study based on a real platform trial for chronic lymphocytic leukemia (presented in supplementary material, Sect.~D).

The remainder of this paper is structured as follows: Section \ref{sect_methods} introduces the notation and considered trial design and describes the proposed methods for platform trials with equal time trends across all arms. In Section \ref{sect_extension}, we discuss a possible extension to trials where the temporal changes affect the arms differently. Section \ref{sect_simstudy} presents an extensive simulation study, evaluating the operating characteristics of the considered methods under a wide range of settings. The use of the modeling strategies is exemplified in two case studies using data on a rare neurodegenerative disease and a type of blood cancer, presented in Section \ref{sect_casestudy}. We conclude the paper with a discussion and conclusions in Section \ref{sect_discussion}.

\section{Methods}\label{sect_methods}

We consider a platform trial with a continuous endpoint that includes $K$ experimental treatments $(K\geq 2)$ and a shared control arm and assume that treatment arms enter the platform trial sequentially. We denote by $k$ the treatment groups where $k=0$ indicates the control and $k=1,\ldots, K$ the experimental treatment arms, indexed by entry order.

\begin{figure}[!h]
    \centering
    \includegraphics[width=0.8\textwidth]{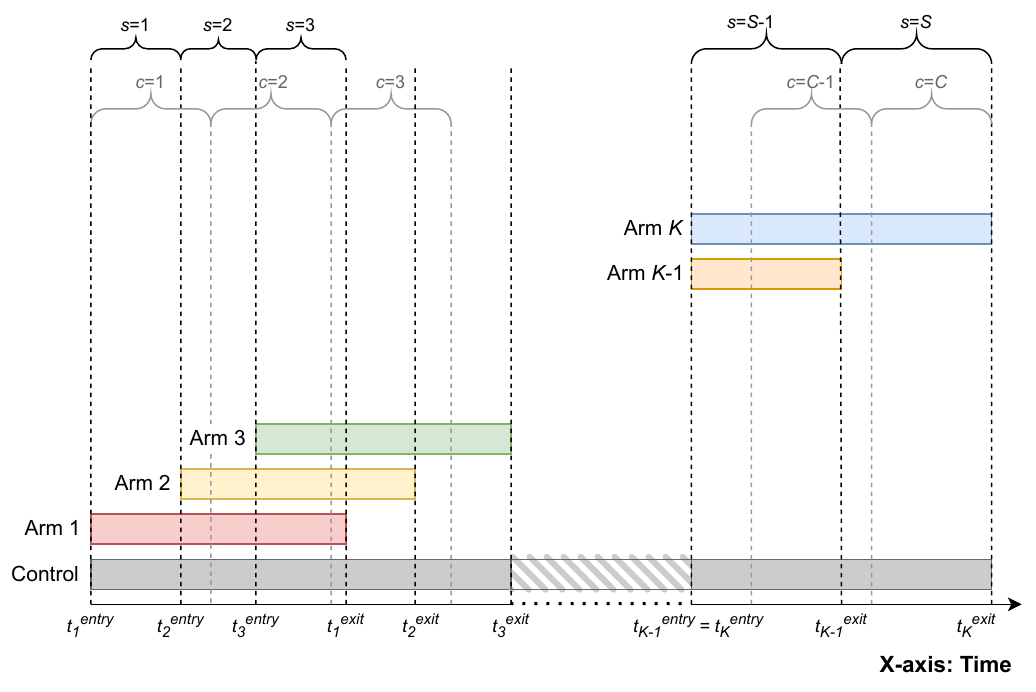}
    \caption{Example of a platform trial with $K$ experimental treatment arms and a shared control group, illustrating the division of the trial duration into periods (denoted by $s$) and units of calendar time (denoted by $c$). The time points $t^{entry}_k$ and $t^{exit}_k$ ($k=1, \ldots, K$) on the x-axis mark the entry and exit times of the experimental treatment arm $k$.}
    \label{fig:trial_scheme}
\end{figure}

We consider two ways of splitting the time in the trial - into periods and calendar time intervals. The periods are defined as time intervals bounded by the time points where experimental treatment arms are added or dropped. Hence, alternations to the total number of arms currently involved in the trial always mark the start of a new period. The periods are indexed by $s=1, \ldots, S$, where $S$ is the total number of periods. The time interval for period $s$ is denoted by $T^S_s$ ($s=1, \ldots, S$). 
An alternative to periods as defined above is to divide the trial duration into $C$ equidistant units of calendar time (e.g. weeks or months), indexed by $c=1, \ldots, C$. We denote the calendar time intervals by $T^C_c$ ($c=1, \ldots, C$) and their length by $c_{length}$. 
Figure \ref{fig:trial_scheme} illustrates the division into periods and calendar time intervals in a platform trial with $K$ experimental treatment arms and a shared control.

We focus on evaluating treatment arms that enter when the trial is already ongoing, and therefore, NCC data is available for these arms. We aim to compare the efficacy of each treatment against the control as soon as the treatment arm is completed, hence, no control data collected after the evaluated arm has finished is used in the analysis.
Consider the one-sided null hypothesis $H_{0,M}:\theta_M \leq 0$ for arm $M$ under study, where $\theta_M$ denotes the treatment effect size for treatment $M$ against control. To test the null hypothesis $H_{0,M}$, we propose several model-based approaches.

To fit the proposed models, we use all data from the trial until treatment $M$ leaves the platform. In other words, we use all data in the set $\mathcal{D}_M = \{ (y_j, k_j, t_j), j=1,...,N \mid t_j \le t^{exit}_M \}$, where $j$ represents the patient index, $y_j$ denotes the response, $k_j$ and $t_j$ are the arm and recruitment time corresponding to patient $j$, respectively, and $t^{exit}_M$ marks the time point when the arm $M$ left the trial. Note that also data from arms for which the recruitment is still ongoing at time $t^{exit}_M$ is used to fit the models. For technical definitions of the time variables, periods, and calendar time intervals, we refer to the supplementary material (see Sect.~A).

In what follows, we introduce the proposed model-based approaches to test the null hypothesis $H_{0,M}$. These models  incorporate the non-concurrent control data. The models adjust for time as a categorical fixed effect, a polynomial spline or a random factor, while stratifying time by either period or calendar time.

\subsection{Fixed effect models}\label{sect_fix}

Firstly, we consider two linear regression models, which estimate the effect of treatments, which were active in the trial prior or up to the time unit $t^{exit}_M$ and adjust for potential time trends by including time as a categorical covariate. The two approaches consider different definitions of the time variable, one using periods and the other one calendar time intervals.

\noindent\textbf{Period adjustment.} 
As previously investigated by \cite{Lee2020Including} and \cite{Bofill2022Model} in simple platform trials with two experimental arms only, in the first model, we adjust for the time effect by including the factor ``period'' to the model, i.e., using a step-function:
\begin{equation}\label{eq_fix_per}
    y_j = \eta_{0,period}  + \sum_{k \in \mathcal{K}_M} \theta_k \cdot I(k_j=k) + \sum_{s=2}^{S_M} \tau_s \cdot I(t_j \in T^{S}_s) + \varepsilon_j
\end{equation}
where $\eta_{0,period}$ is the response in the control arm in the first period; $\theta_k$ represents the effect of treatment $k$ compared to control for $k\in\mathcal{K}_M$, where $\mathcal{K}_M$ 
is the set of treatments that were active in the trial during periods prior or up to $t^{exit}_M$; $\tau_s$ indicates the stepwise period effect between periods 1 and $s$ ($s = 2, \ldots, S_M$), where $S_M = \{s \mid t^{exit}_M \in T^S_s\}$ denotes the period, in which arm $M$ left the trial (i.e. the period in which $t^{exit}_M$ is included). 

\vspace{0.3cm}
\noindent\textbf{Calendar time adjustment.} In the second approach, analogously to the Bayesian Time Machine \citep{Saville2022Bayesian}, we employ calendar time intervals to adjust for time trends. Specifically, we consider a regression model adjusting for the time effect by using calendar time intervals, and thus estimating the calendar time effect rather than period effect:
\begin{equation}\label{eq_fix_cal}
    y_j = \eta_{0,calendar}  + \sum_{k \in \mathcal{K}_M} \theta_k \cdot I(k_j=k) + \sum_{c=2}^{C_M} \tau_c \cdot I(t_j \in T^{C}_c) + \varepsilon_j
\end{equation}
Here $\eta_{0,calendar}$ represents the control response in the first calendar time unit; $\theta_k$ denotes the treatment effect of treatment $k$ compared to control for $k \in \mathcal{K}_M$; $\tau_c$ is the stepwise effect between calendar time units 1 and $c$ ($c = 2, \ldots, C_M$), where $C_M$ indicates the calendar time unit, in which arm $M$ left the trial. The length of the calendar time units $c_{length}$ is given in terms of the number of enrolled patients and can be specified arbitrarily. Note that since only data until $t_M^{exit}$ are used to fit the model, the last calendar time unit $C_M$ will typically be shorter than $c_{length}$, as entering and leaving times of the experimental treatment arms are independent of the calendar time unit length $c_{length}$.

Both models \eqref{eq_fix_per} and \eqref{eq_fix_cal} implicitly assume a time trend that is constant in all periods or calendar time intervals, respectively, and that the time effect is equal across all arms and additive on the model scale. In both models, the residuals $\varepsilon_j$ are assumed to follow a normal distribution $\varepsilon_j \sim \mathcal{N}(0, \sigma^2)$. Moreover, the length of the time intervals in \eqref{eq_fix_cal} poses an additional design parameter that allows to adjust for longer or shorter time intervals than those given by the periods, which are also independent of the alternations to the trial design.

\subsection{Spline regression}\label{splines}

The models with a categorical fixed effect \eqref{eq_fix_per} and \eqref{eq_fix_cal} assume that the time trend is constant within each period or calendar time interval. In order to also capture more complex patterns of the time trend, we consider estimating the patient response over time using spline regression. The model is given by:
\begin{equation}\label{eq_spline_reg}
    y_{j} = \Tilde{\eta}_0  + \sum_{k \in \mathcal{K}_M} \theta_k \cdot I(k_j=k) + f(t_j) + \varepsilon_{j}
\end{equation}
where $y_j$, and $\theta_k$ are defined as in \eqref{eq_fix_per} and $\Tilde{\eta}_0$ denotes the model intercept. The residuals are $\varepsilon_j \sim \mathcal{N}(0, \sigma^2)$. Note that the treatment effect enters the model as a linear predictor. The time trend is modeled via a continuous function $f(t_j)$ of the patient entry time $t_j$.

Specifically, we consider the B-spline function to model the time trend. This function is composed of multiple polynomial functions of a given degree $q$, which are joined together at points called \textit{knots}, such that the entire spline is continuously differentiable up to the $(q-1)$th derivative \citep{Eilers1996Flexible}. For the positions of these knots, we evaluate two strategies, somewhat analogous to the period and calendar time adjustments. Firstly, we place the inner knots to the beginning of each period $s=2,\ldots,S$, such that one polynomial of degree $q$ is always fitted to each period. Moreover, we consider placing the inner knots equidistantly, according to the length of the calendar time unit, and thus fitting a polynomial of degree $q$ to each calendar time interval. Regarding the degree of the fitted polynomial, we explore linear, quadratic and cubic splines, i.e. $q \in (1,2,3)$. For details on the definition of the B-spline, see Sect.~B.1 in the supplementary material.

Modeling the time using spline functions gives the model additional flexibility compared to the regression model that adjusts for time using a linear function considered in \cite{Lee2020Including} and \cite{Bofill2022Model}, as it allows for more accurate modeling of complex time trend patterns.

\subsection{Mixed effect models}\label{sect_mix}

In models \eqref{eq_fix_per} and \eqref{eq_fix_cal}, time is considered as a fixed factor. Alternatively, patients within different periods or calendar time units could be considered as different clusters, having a period- or calendar time-specific random intercepts. In what follows, we include the time variable to the models as a random factor. Under such models, the potential correlation between the random effects associated with different periods or calendar times can also be taken into account.

\subsubsection{Mixed models with uncorrelated random effects.} First, we consider simple mixed effect models, where the effects of the given time intervals (period or calendar time units) are assumed to be uncorrelated with the effects of neighboring intervals. The mixed effect model with period adjustment has the following form:
\begin{equation} \label{eq_mix_per}
    y_{j} = \gamma_{0,period}  + \sum_{k \in \mathcal{K}_M} \theta_k \cdot I(k_j=k) + \sum_{s=2}^{S_M} u_s \cdot I(t_j \in T^{S}_s) + \varepsilon_{j}
\end{equation}
whereas the model adjusting for calendar time units is given by:
\begin{equation} \label{eq_mix_cal}
    y_{j} = \gamma_{0,calendar}  + \sum_{k \in \mathcal{K}_M} \theta_k \cdot I(k_j=k) + \sum_{c=2}^{C_M} u_c \cdot I(t_j \in T^{C}_c) + \varepsilon_{j}
\end{equation}
where $y_{j}$ and $\theta_k$ have the same interpretation as in the fixed effect models. The model intercepts $\gamma_{0,period}$ and $\gamma_{0,calendar}$ are, in this case, given in terms of the control response across the whole trial up until $S_M$ or $C_M$, respectively. $u_s$ and $u_c$ denote the random effect associated with the intercept for period $s$ or calendar time unit $c$. These period- and calendar time-specific random effects are assumed to be normally distributed with mean 0 and constant variances $\sigma^2_{period}$ and $\sigma^2_{calendar}$, respectively: $u_s \sim \mathcal{N}(0, \sigma^2_{period})$ and $u_c \sim \mathcal{N}(0, \sigma^2_{calendar})$. Note that in this case, the correlation between any two period or calendar time effects is 0.

In both models, the distribution of the residuals $\varepsilon_{j}$, associated with the response of individual patient $j$ is assumed to be the same for all treatments: $\varepsilon_j \sim \mathcal{N}(0, \sigma^2)$.

\subsubsection{Mixed models with autocorrelated random effects.} To account for possible correlation of the random effects, we also consider random effects with first-order autoregressive structure (AR(1)), again with period or calendar time adjustments. The model equations are identical to \eqref{eq_mix_per} and \eqref{eq_mix_cal}.

There is a difference, however, with respect to the distribution of the random effects, which are now modeled as autocorrelated. Hence, the random effects for individual periods and calendar time units are assumed to be normally distributed with mean 0, constant variance $\sigma^2_{period}$ or $\sigma^2_{calendar}$, and an AR(1) correlation structure. For details, we refer to the supplementary material (see Sect.~B.2).
The model residuals $\varepsilon_j$ are also assumed to be normally distributed (i.i.d.) with mean 0 and constant variance $\sigma^2$.

\section{Extension to relax the assumption of equal time trends}\label{sect_extension}

The modeling approaches outlined in the previous section, as well as all other modeling approaches available to date for utilizing non-concurrent controls, rely on the assumption of equal time trends in all arms in the platform trial on the model scale \citep{Bofill2022Commentary}. Violation of this assumption can lead to the loss of type I error rate control and biased treatment effect estimates. In this section, we propose an analysis method that could mitigate potential biases in trials where there are different time trends across arms.

In particular, we extend the mixed models from Section \ref{sect_mix} and consider mixed models with treatment and time as categorical fixed effects (using the period definition of the time variable, as well as calendar time intervals) and the interaction between these variables as a random effect. The model adjusting for periods is defined  as follows:
\begin{equation}\label{eq_mix_int_per}
\begin{split}
    y_j & = \eta_{0,period}  + \sum_{k \in \mathcal{K}_M} \theta_k \cdot I(k_j=k) + \sum_{s=2}^{S_M} \tau_s \cdot I(t_j \in T^{S}_s) + \\
    & + \sum_{k \in \mathcal{K}_M \setminus M} \sum_{s=2}^{S_M} u_{k,s} \cdot I(k_j=k) \cdot I(t_j \in T^{S}_s)  + \varepsilon_j
\end{split}
\end{equation}
Similarly, the model with calendar time adjustment is specified as:
\begin{equation}\label{eq_mix_int_cal}
\begin{split}
    y_j & = \eta_{0,calendar}  + \sum_{k \in \mathcal{K}_M} \theta_k \cdot I(k_j=k) + \sum_{c=2}^{C_M} \tau_c \cdot I(t_j \in T^{C}_c) + \\
    & + \sum_{k \in \mathcal{K}_M \setminus M} \sum_{c=2}^{C_M} u_{k,c} \cdot I(k_j=k) \cdot I(t_j \in T^{C}_c)  + \varepsilon_j
\end{split}
\end{equation}

For the terms $y_j$, $\eta_{0,period}$, $\eta_{0,calendar}$, $\theta_k$, $\tau_s$ and $\tau_c$, the same interpretation as in the fixed effect models described in Section \ref{sect_fix} holds. The interaction term between treatment arm and time interval is included as a random factor. Hence, the effects associated with treatment arm $k~(k\neq M)$ and period $s$ (or treatment arm $k~(k\neq M)$ and calendar time unit $c$), denoted by $u_{k,s}$ and $u_{k,c}$, respectively, are drawn from a normal distribution with mean 0 and constant variances: $u_{k,s} \sim \mathcal{N}(0, \sigma^2_{arm,period})$ and $u_{k,c} \sim \mathcal{N}(0, \sigma^2_{arm,calendar})$ and are modeled as uncorrelated. Since it is assumed that the time trends in the evaluated treatment arm $M$ and the control are equal, the arm $M$ is excluded from the interaction term. As before, $\varepsilon_j \sim \mathcal{N}(0, \sigma^2)$ is assumed for the model residuals.

These models estimate the differences in the average response in each treatment arm relative to the control and each period/calendar time unit relative to the first one. Moreover, the interaction terms allow for a random variation in the response for each treatment arm in each period/calendar time unit. As the random effects are centered around 0, this approach is shrinking this random variation towards 0. While under different time trends, this approach may not completely remove the bias (due to the shrinkage), it is expected to reduce the bias compared to a model without the interaction term.

\section{Simulation study}\label{sect_simstudy}

We evaluate the properties of the methods proposed in Sections \ref{sect_methods} and \ref{sect_extension} in a simulation study and discuss the influence of certain design parameters on the type I error rate and statistical power.

\subsection{Design}

We consider a platform trial with $K$ experimental treatment arms that enter the trial sequentially and a control group that is common to all treatment arms. In the simulations, we assume that the recruitment is uniform with exactly one patient being enrolled at each time point. The timing of adding of the treatment arms can then be expressed in terms of the cumulative sample size, given by $\mathbf{d} = (d_1, \ldots, d_K)$, where $d_k$ indicates how many patients had already been enrolled to the trial by the time treatment $k$ entered the platform. Note that in this special case of uniform patient recruitment, we have $t_k^{entry}=d_k$. We always set the $d_1$ to 0 to ensure that the platform trial starts with at least one experimental treatment. In this work, we consider platform trials with equidistant entry times of the experimental treatment arms, where the arm $k$ enters the trial after $d_k = d \cdot (k-1)$ patients have been recruited to the platform.
Note that the parameter $d$ determines the amount of overlapping sample size between the treatment arms. If $d=0$, all arms join and leave the trial simultaneously, resulting in a standard multi-arm trial and a total overlap between the arms. If $d=2n$, a new treatment arm enters the trial once the previous one finishes, so that there is only one active experimental arm at a time, and hence no overlap between them. 

\subsubsection{Data generation}\label{sect_data_gen}

To generate the trial data we assume an allocation ratio of $1:1: \ldots :1$ in each period. Patients are assigned to arms following block randomization with block sizes of $2 \cdot (\text{number of active arms} +1)$ at each time $t$. Patients are indexed by entry order, assuming that at each time unit exactly one patient is recruited (i.e., $t_j = j$) and the time of recruitment and of the observation of the response are equal. The continuous outcome $y_j$ for patient $j$ is then drawn from a normal distribution according to $y_j \sim \mathcal{N}(\mu, \sigma^2)$ with 
\begin{gather*}
    \mu = \eta_0 + \sum_{k=1}^K \theta_k \cdot I(k_j = k) + f(j) \text{ and } \sigma^2 = 1
\end{gather*}

where $\eta_0$ and $\theta_k$ are the response in the control arm and the effect of treatment $k$. Moreover, time trends of various strengths and shapes may be present in the trial. The time trends are denoted by the function $f(j)$ and their magnitude is arm $k$ is parameterized by $\lambda_k$. The following time trend patterns are considered:

\begin{itemize}
    \item[] Inverted-U time trend: $f(j) = \lambda_k \cdot \frac{j-1}{N-1} \text{ for } j \leq N_p$ and $f(j) = -\lambda_k \cdot \frac{j-N_p}{N-1} + \lambda_k \cdot \frac{N_p-1}{N-1} \text{ for } j > N_p$, where $N$ indicates the total sample size in the trial and $N_p$ is the point at which the trend turns from positive to negative in terms of the sample size. For $\lambda_k>0$ the mean response in arm $k$ linearly increases (with slope $\lambda_k$) until the total sample size has reached $N_p$, and linearly decreases afterwards. For $\lambda_k<0$, it decreases until $N_p$ and increases afterwards.

    \item[] Linear time trend: $f(j) = \lambda_k \cdot \frac{j-1}{N-1}$, where $N$ indicates the total sample size in the trial. The mean response in arm $k$ linearly increases with the slope $\lambda_k$ over time.

    \item[] Seasonal time trend: $f(j) = \lambda_k \cdot \mathrm{sin} \big( \psi \cdot 2\pi \cdot \frac{j-1}{N-1} \big)$, where $N$ indicates the total sample size in the trial. The seasonal trend may consist of multiple cycles, where the response increases at first and decreases afterwards, while the respective peaks of the cycles correspond to $\lambda_k$ or $-\lambda_k$. The number of cycles over the whole platform trial is determined by $\psi$.
    
    \item[] Stepwise time trend: $f(j) = \lambda_k \cdot (i_j - 1)$, where $i_j$ indicates how many treatment arms have already entered the trial at the time patient $j$ was enrolled. Hence, there is a jump in the mean response in arm $k$ of size $\lambda_k$ every time a new arm is added to the trial. If multiple arms are added simultaneously, the jump in the response is multiplied by how many arms have entered. Note that the stepwise time trend is in general stronger than the other patterns and hence not directly comparable to them. This is in particular pronounced in trials with many treatment arms, as the trend increases by $\lambda_k$ every time a new arm enters the platform. The rationale behind considering this pattern in the simulations is to examine the behavior of the models in more extreme situations.
\end{itemize}

\begin{figure}[!h]
    \centering
    \includegraphics[width=\textwidth]{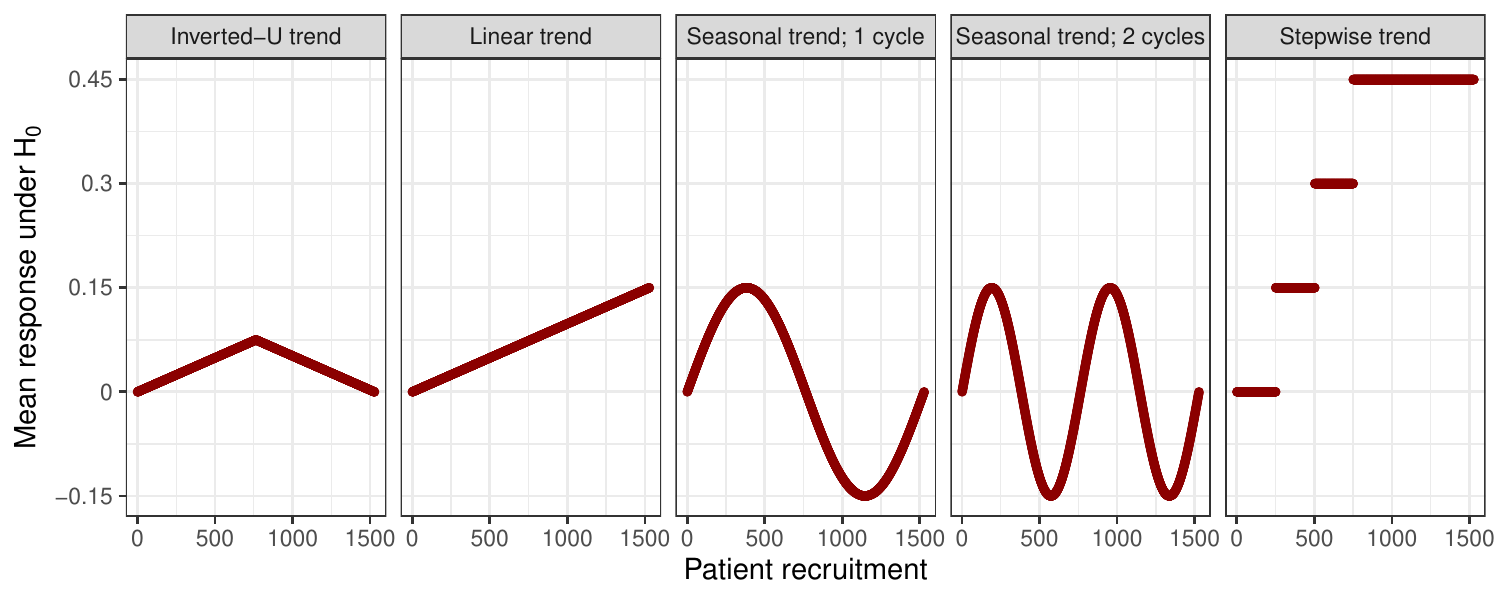}
    \caption{Mean responses under the null hypothesis under time trends of different patterns and strength of $\lambda_k=0.15$ $\forall k$ in a platform trial with 4 experimental treatment arms entering the trial sequentially.}
    \label{fig:trend_patterns}
\end{figure}

The considered time trend patterns are illustrated in Figure \ref{fig:trend_patterns}.

\subsubsection{Considered trial settings}

We distinguish three settings in which we vary the design according to the objectives of the simulation study:

\noindent \textbf{Setting 1:} With the objective of assessing the impact of the overlap between arms, we focus on a larger platform trial with $K=10$ experimental arms and equal time trend in all arms ($\lambda_k = \lambda_0$ $\forall k$). We compare arm $M=5$ to the control, while varying the strength of the time trend ($\lambda_k$).
    \begin{itemize}
        \item[] Setting 1A: Focusing on trials with a linear time trend, here, we additionally vary the amount of overlap between the treatment arms by varying the entry times. In this setting, we evaluate the impact of the overlapping sample size on the performance of the fixed effect regression model with period adjustment \eqref{eq_fix_per}. 
        
        \item[] Setting 1B: Here, we additionally vary the pattern of the time trend, considering all time trend shapes listed above. In this setting, we assess the properties of the spline regression \eqref{eq_spline_reg} under different time trend patterns with varying strengths. 
    \end{itemize}

\noindent \textbf{Setting 2:} In order to understand the behavior of the fixed effect model using the alternative definition of time, as well as the mixed effect models, we investigate a simpler platform trial with $K=4$ experimental arms and equal time trend in all arms ($\lambda_k = \lambda_0$ $\forall k$). We compare arm $M=3$ to the control, while varying the pattern (considering all patterns listed above) and strength of the time trend ($\lambda_k$).
    \begin{itemize}
        \item[] Setting 2A: Here, we additionally vary the size of the calendar time unit, in order to examine the influence of this parameter on the operating characteristics of the fixed effect model adjusting for calendar time \eqref{eq_fix_cal}. We compare the performance of this model to the fixed effect model with period adjustment \eqref{eq_fix_per}. 
        
        \item[] Setting 2B: Using calendar time unit size of 100 patients, in this setting we evaluate the performance of the mixed effect models without interaction (\eqref{eq_mix_per} and \eqref{eq_mix_cal}). Considering different time trend patterns with varying strength, we compare these models to the fixed effect model adjusting for periods \eqref{eq_fix_per}.
    \end{itemize}

\noindent \textbf{Setting 3:} Keeping the platform trial with $K=4$ experimental arms from the previous setting, we focus on examining trials with different time trends across arms. Again, we compare arm $M=3$ to the control, while varying the pattern (considering all patterns listed above) and strength of the time trend ($\lambda_k$). In particular, the treatment arm 3 and the control group have no time trend ($\lambda_0 = \lambda_3 = 0$), while the rest of the arms have time trends of different strengths. We vary the strength of the time trend in arms 1 ($\lambda_0 = \lambda_2 = \lambda_3 = \lambda_4 \neq \lambda_1$), 1 and 2 ($\lambda_0 = \lambda_3 = \lambda_4 \neq \lambda_1 = \lambda_2$), or 1, 2 and 4 ($\lambda_0 = \lambda_3 \neq \lambda_1 = \lambda_2 = \lambda_4$ and $\lambda_0 = \lambda_3 \neq \lambda_1$; $\lambda_2 = 2 \lambda_1$; $\lambda_4 = 3 \lambda_1$). In this setting, we evaluate the performance of the mixed effect models with interaction (\eqref{eq_mix_int_per} and \eqref{eq_mix_int_cal}) and compare it to the fixed effect model adjusting for periods \eqref{eq_fix_per}. Note that we only consider cases where for the evaluated treatment arm, the time trend is equal to the one in the control group. If this assumption does not hold, the considered testing approaches would test the null hypothesis that the average treatment effect over the time when the treatment is active in the platform is lower or equal to 0. This would imply that the treatment effect depends on the time periods, in which the treatment is active in the platform.

\vspace{0.5cm}

In all cases, we assume an underlying mean response of zero for the control arm ($\eta_0 = 0$) and a treatment effect of $\theta_k=0.25$ for treatment arms under the alternative hypothesis, as well as sample sizes of $n=250$ for all experimental treatment arms. This treatment effect was chosen such that the analysis using CC data only achieves approximately 80\% power with the given sample size at a one-sided significance level $\alpha=0.025$. 

For comparative purposes, we also analyze the data with two simple standard approaches, where the data from the investigated treatment arm $M$ is compared either to pooled NCC and CC data (pooled analysis) or only CC data (separate analysis) with one-sided t-tests \citep{Viele2014Use}. Moreover, we include the fixed effects model with period adjustment \eqref{eq_fix_per} as a reference model for all comparisons.

Table \ref{tab:tab_scenarios} summarizes the considered simulation settings and parameters. We simulated 10,000 replicates of each scenario to estimate the type I error rate and statistical power.

\begin{landscape}

\begin{table}[]
\resizebox{0.98\paperwidth}{!}{\begin{tabular}{|c|c|c|c|c|c|c|c|c|}
\hline
\textbf{Setting} & $K$ & $d$ & $\lambda$ & \begin{tabular}[c]{@{}c@{}}\textbf{Trend}\\ \textbf{pattern}\end{tabular} & \begin{tabular}[c]{@{}c@{}}\textbf{Calendar}\\ \textbf{time unit}\\ \textbf{size}\end{tabular} & \textbf{Objective} & \begin{tabular}[c]{@{}c@{}}\textbf{Main findings in the}\\ \textbf{considered settings}\end{tabular} & \begin{tabular}[c]{@{}c@{}}\textbf{Conditions}\\ \textbf{for T1E control}\end{tabular} \\ \hline
\textbf{1A} & 10 & \begin{tabular}[c]{@{}c@{}}$d_k = d \cdot (k-1)$ $\forall k \ge 1$ \\ - $d \in [0, 500]$ \\ - with increments of 25\end{tabular} & \begin{tabular}[c]{@{}c@{}}$\lambda_k = \lambda_0 \in [-0.5, 0.5]$ $\forall k$\\ - with increments of 0.125\end{tabular} & Linear & - & \begin{tabular}[c]{@{}c@{}}Evaluate the fixed effect regression \\ model with period adjustment \eqref{eq_fix_per}\end{tabular} & \begin{tabular}[c]{@{}c@{}} The fixed effect model with period adjustment \\ leads to power gains compared to the separate \\ analysis, provided that there is some overlap \\ between the treatment arms.\end{tabular}  & \begin{tabular}[c]{@{}c@{}}Equal and additive \\ time trends, block \\ randomization. \end{tabular} \\ \hline
\textbf{1B} & 10 & \begin{tabular}[c]{@{}c@{}}$d_k = d \cdot (k-1)$ $\forall k \ge 1$ \\ - $d = 250$\end{tabular} & \begin{tabular}[c]{@{}c@{}}$\lambda_k = \lambda_0 \in [-0.5, 0.5]$ $\forall k$\\ - with increments of 0.125\end{tabular} & \begin{tabular}[c]{@{}c@{}}Linear,\\ stepwise,\\ inverted-U\\ seasonal\end{tabular} & 450 & \begin{tabular}[c]{@{}c@{}}Evaluate the spline\\ regression \eqref{eq_spline_reg}\end{tabular} & \begin{tabular}[c]{@{}c@{}}Spline regression controls the type I error rate \\ if the time trend pattern is sufficiently smooth \\ and equal across all arms. In cases with little \\ to no overlap between the treatment arms, \\ splines achieve larger power than the regression \\ model with period adjustment.\end{tabular}  & \begin{tabular}[c]{@{}c@{}}Equal and additive \\ time trends, sufficiently \\ smooth trend pattern, \\ block randomization. \end{tabular} \\ \hline
\textbf{2A} & 4 & \begin{tabular}[c]{@{}c@{}}$d_k = d \cdot (k-1)$ $\forall k \ge 1$ \\ - $d = 250$\end{tabular} & \begin{tabular}[c]{@{}c@{}}$\lambda_k = \lambda_0 \in [-0.5, 0.5]$ $\forall k$\\ - with increments of 0.125\end{tabular} & \begin{tabular}[c]{@{}c@{}}Linear,\\ stepwise,\\ inverted-U\\ seasonal\end{tabular} & \begin{tabular}[c]{@{}c@{}}{[}25, 750{]}\\ with\\ increments\\ of 25\end{tabular} & \begin{tabular}[c]{@{}c@{}}Evaluate the definition of\\ time as calendar time intervals\\  in the fixed effect model \eqref{eq_fix_cal}\end{tabular} & \begin{tabular}[c]{@{}c@{}}Type I error rate inflation depends \\ on the unit size, and pattern and strength \\ of the time trend. Exact type I error rate \\ control is only achieved if calendar time \\ intervals match the entry and leaving times \\ of the experimental arms and the time trends \\ are equal across arms. Larger calendar \\ time units lead to larger power gains.\end{tabular}  & \begin{tabular}[c]{@{}c@{}}Equal and additive \\ time trends, appropriate \\ cal. time unit size, \\ block randomization. \end{tabular} \\ \hline
\textbf{2B} & 4 & \begin{tabular}[c]{@{}c@{}}$d_k = d \cdot (k-1)$ $\forall k \ge 1$ \\ - $d = 250$\end{tabular} & \begin{tabular}[c]{@{}c@{}}$\lambda_k = \lambda_0 \in [-0.5, 0.5]$ $\forall k$\\ - with increments of 0.125\end{tabular} & \begin{tabular}[c]{@{}c@{}}Linear,\\ stepwise,\\ inverted-U\\ seasonal\end{tabular} & 100 & \begin{tabular}[c]{@{}c@{}}Evaluate the mixed models\\ without interaction \eqref{eq_mix_per}, \eqref{eq_mix_cal}\end{tabular} & \begin{tabular}[c]{@{}c@{}}Mixed models with period as random factor \\ do not control the type I error rate in the \\ presence of time trends. Type I error rate \\ inflation is more pronounced in the case \\ of moderate time trends.\end{tabular} & \begin{tabular}[c]{@{}c@{}} Type I error rate is not \\ controlled under positive \\ time trends and conservative \\ under negative time trends. \end{tabular} \\ \hline
\textbf{3} & 4 & \begin{tabular}[c]{@{}c@{}}$d_k = d \cdot (k-1)$ $\forall k \ge 1$ \\ - $d = 250$\end{tabular} & \begin{tabular}[c]{@{}c@{}}\textbf{1)} $\lambda_1 \in [-0.5, 0.5]$;\\ $\lambda_0 = \lambda_2 = \lambda_3 = \lambda_4 = 0$\\ \\ \textbf{2)} $\lambda_1 = \lambda_2 \in [-0.5, 0.5]$;\\ $\lambda_0 = \lambda_3 = \lambda_4 = 0$\\ \\ \textbf{3)} $\lambda_1 = \lambda_2 = \lambda_4 \in [-0.5, 0.5]$;\\ $\lambda_0 = \lambda_3 = 0$\\ \\ \textbf{4)} $\lambda_1 \in [-0.5, 0.5]$ \\ $\lambda_2 = 2 \lambda_1$ \\ $\lambda_4 = 3 \lambda_1$\\ $\lambda_0 = \lambda_3 = 0$ \\ \\ \textbf{5)} $\lambda_k = \lambda_0 \in [-0.5, 0.5]$ $\forall k$\\ \\ - intervals with increments of 0.125\end{tabular} & Linear & 100 & \begin{tabular}[c]{@{}c@{}}Evaluate the mixed models\\ with interaction \eqref{eq_mix_int_per}, \eqref{eq_mix_int_cal}\end{tabular} & \begin{tabular}[c]{@{}c@{}}Mixed models with fixed effect period and \\ the interaction between period and treatment \\ as random effect reduce the type I error rate \\ inflation in settings with unequal time trends \\ across arms compared to the fixed effect model \\ with period adjustment, but do not eliminate \\ it completely.\end{tabular}  & \begin{tabular}[c]{@{}c@{}}Equal and additive \\ time trends, appropriate \\ cal. time unit size, \\ block randomization. \\ Under unequal time trends \\ smaller T1E inflation than \\ fixed effect model \eqref{eq_fix_per}. \end{tabular} \\ \hline
\end{tabular}}
\caption{Simulation settings and parameters considered in the simulation study, and main findings for each considered scenario.}
\label{tab:tab_scenarios}
\end{table}

\end{landscape}

\subsubsection{Implementation}

The simulation study was performed using the \texttt{NCC} R package \citep{NCC_pkg}, using the version labeled as Release 1.4 on GitHub. The package contains functions implementing all considered analysis approaches, as well as functions for data simulation and performing simulation studies \citep{krotka2023ncc}.

\subsection{Results}

For better legibility and easier comparison, we restricted the range of the y-axis in all plots showing the statistical power to [0.7, 1]. If the power is not visible for some values on the x-axis, the estimated power was below 0.7. In all plots showing the estimated type I error rate, we include a dashed reference line for the nominal significance level of 0.025 and a grey area around this value representing the 95\% prediction interval of the simulated type I error rate with 10,000 simulation runs, provided that the true type I error rate is 0.025.

\subsubsection{Setting 1: Assessing the properties of the fixed effect model with period adjustment and of the spline regression}

Considering a platform trial with 10 experimental treatment arms and a shared control arm, we assess the operating characteristics of the fixed effect model adjusting for periods and the spline regression. In particular, we evaluate the effect of the overlapping sample size and the strength and pattern of the time trend on the type I error rate and statistical power.

\begin{figure}[h!]
    \centering
    \includegraphics[width=1\textwidth]{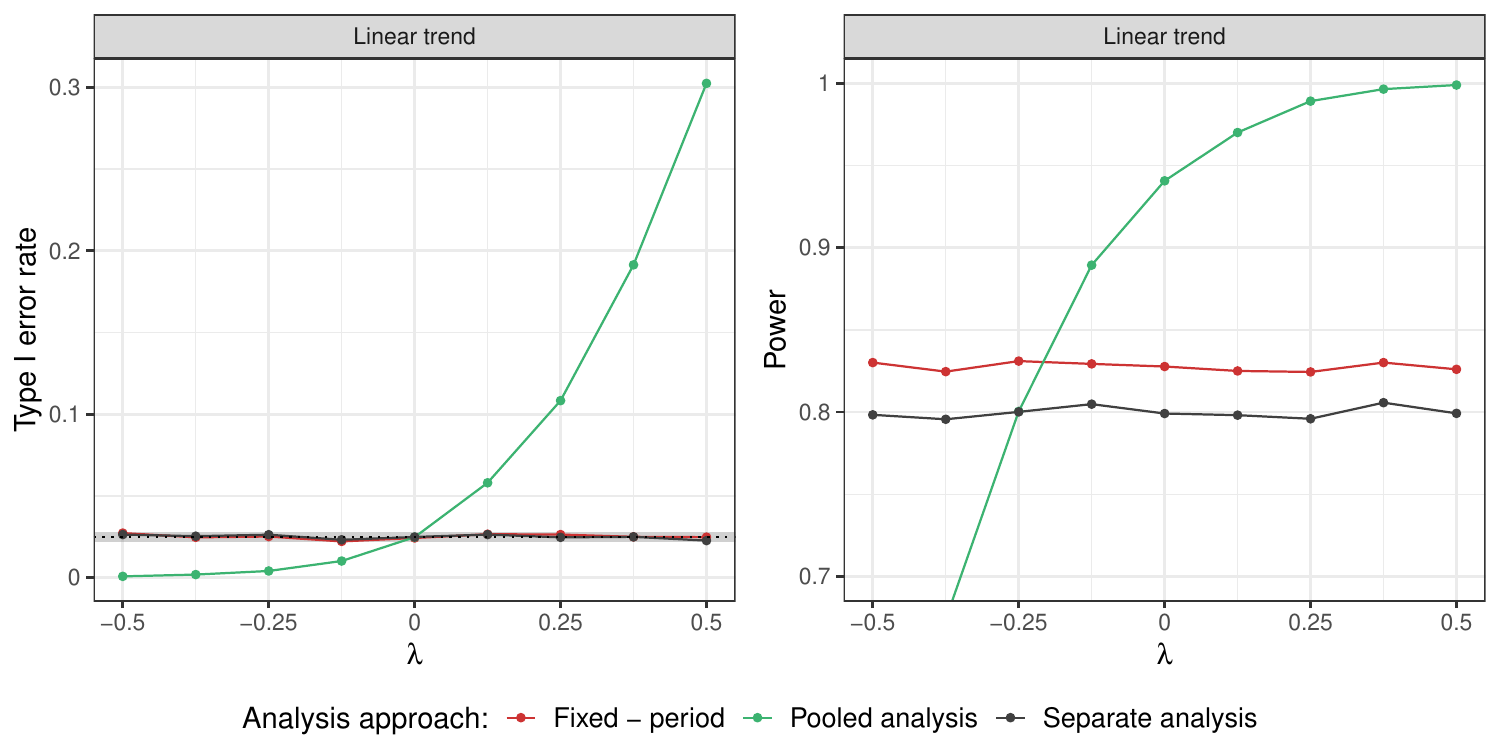}
    \caption{\textbf{Setting 1A:} Type I error rate and power of the fixed effect regression model with period adjustment compared to the pooled and separate analyses with respect to the strength of the time trend $\lambda$. Here, $d=400$, $n=250$ in each experimental arm, and a linear shape of the time trend are used. Treatment arm 5 is being evaluated.}
    \label{fig:fixmodel_lambda}
\end{figure}

\textbf{Setting 1A.} Figure \ref{fig:fixmodel_lambda} shows the impact of the strength of the time trend on the operating characteristics when evaluating treatment arm 5.
Analogous plots for all treatment arms can be found in Figure S1 in the supplementary material. 
The regression model and the separate analysis asymptotically control the type I error rate, regardless of the strength of the time trend. The pooled analysis leads to inflation of the type I error in the presence of positive time trends and deflation in case of negative time trends. Additionally, the model leads to gains in power as compared to the separate analysis. This result is in line with previous works \citep{Lee2020Including, Bofill2022Model}, where the regression model was investigated for trials with two experimental arms only.

We vary the overlap between arms by varying the parameter $d$ from 0 to $2n=500$. The effect of the amount of overlapping sample size when evaluating treatment arm 5 is shown in Figure \ref{fig:fixmodel_d}. Figure S2 in the supplementary material shows analogous plots for all treatment arms. 
The overlaps do not affect the type I error rate control of the regression model and separate approach, which is guaranteed (asymptotically) in all the cases. The inflation of the pooled analysis gets stronger with increasing $d$. This is because larger values of $d$ result in longer platform trials and larger size of the NCC data. The power of the regression model, however, depends on the overlap between treatment arms. In the extreme cases with $d=0$ and $d=2n$, the regression model leads to identical power as the separate analysis. If $d=0$, there is no NCC data, as all the arms join the trial at the beginning. Thus, the control group used for the treatment-control comparison is the same for both, the separate analysis and the regression model. In case of no overlap between the arms ($d=2n=500$), there is insufficient amount of data to estimate the period effect. Hence, simultaneous presence of the experimental arms in the trial is crucial for a reliable estimation of the period effect and resulting power gains when using the regression model. For the considered simulations, the maximal power is reached for $d=175$, which in this case leads to the optimal trade-off between the amount of overlapping sample size between treatment arms and the size of the non-concurrent data.

\begin{figure}[h!]
    \centering
    \includegraphics[width=1\textwidth]{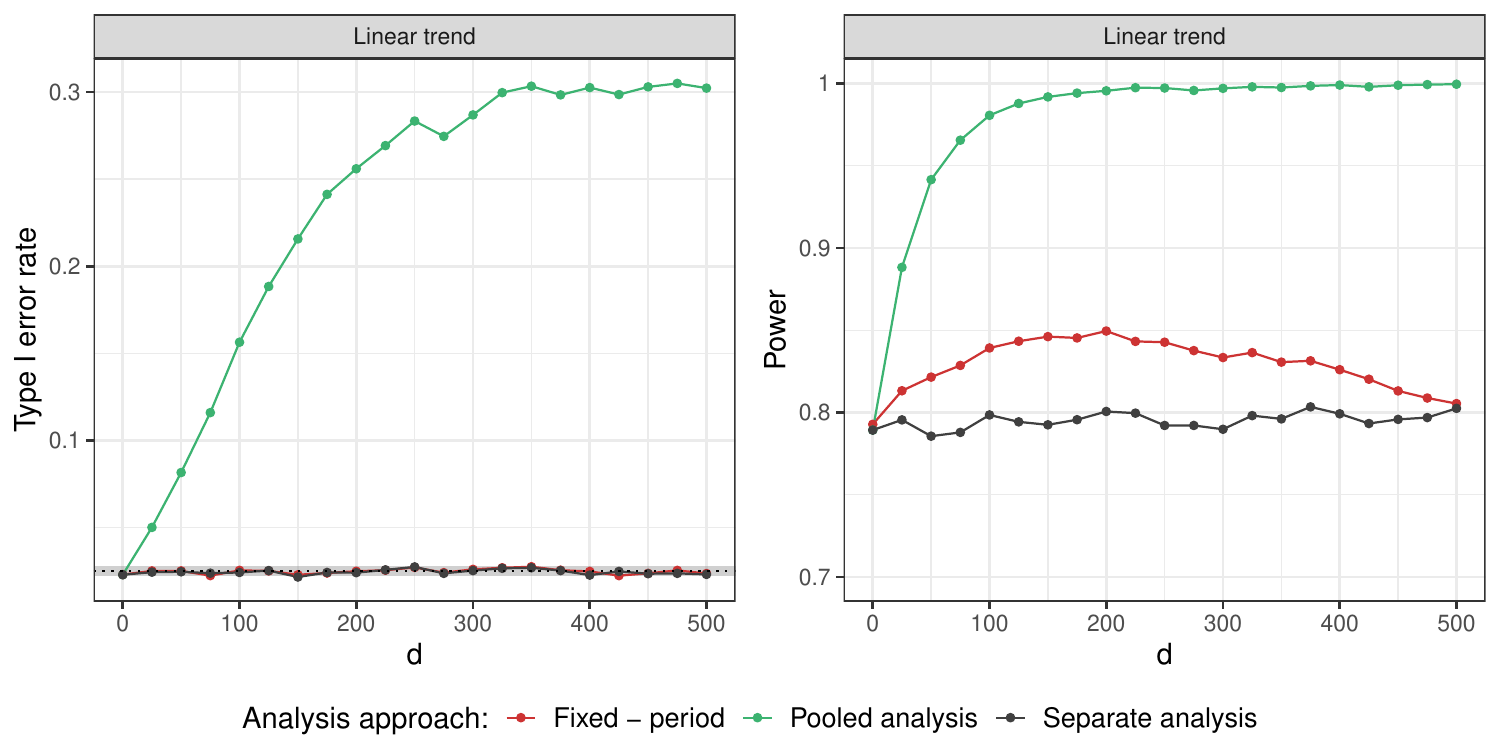}
    \caption{\textbf{Setting 1A:} Type I error rate and power of the fixed effect regression model with period adjustment compared to the pooled and separate analyses with respect to the timing of adding the treatment arms. Here, $n=250$ in each experimental arm and a linear time trend with strength $\lambda=0.5$ are considered. Treatment arm 5 is being evaluated.}
    \label{fig:fixmodel_d}
\end{figure}

Figure S3 in the supplementary material illustrates how the type I error rate and power for individual treatment-control comparisons depends on the order of entry in the platform trial. We observe that the power of the regression model increases for arms that were added to the trial later. This is due to larger sample size of the NCC data. On the other hand, this also leads to a higher inflation of the type I error with the pooled analysis, while the regression model controls the type I error for all treatment-control comparisons.

To sum up, the fixed effect regression model with period adjustment offers superior performance in terms of the statistical power in trials where the treatment arms enter in a staggered way but overlap for some time. Arms added to the trial later have the advantage of more NCC data available for the treatment-control comparison, hence there is higher probability to identify efficacious experimental arms as such.

\textbf{Setting 1B.} The performance of the spline regression was evaluated under varying overlap between the treatment arms, as well as time trends of different patterns and strengths. We also varied the degree of the B-splines in the simulations, considering linear, quadratic, and cubic splines. However, since the differences in the resulting operating characteristics were only marginal, here we only present results for the cubic spline regression. 
Results for linear and quadratic splines can be found in supplementary Figures S4 and S5. The knots are placed either at the beginning of each period or each calendar time interval.

\begin{figure}[h!]
    \centering
    \includegraphics[width=1\textwidth]{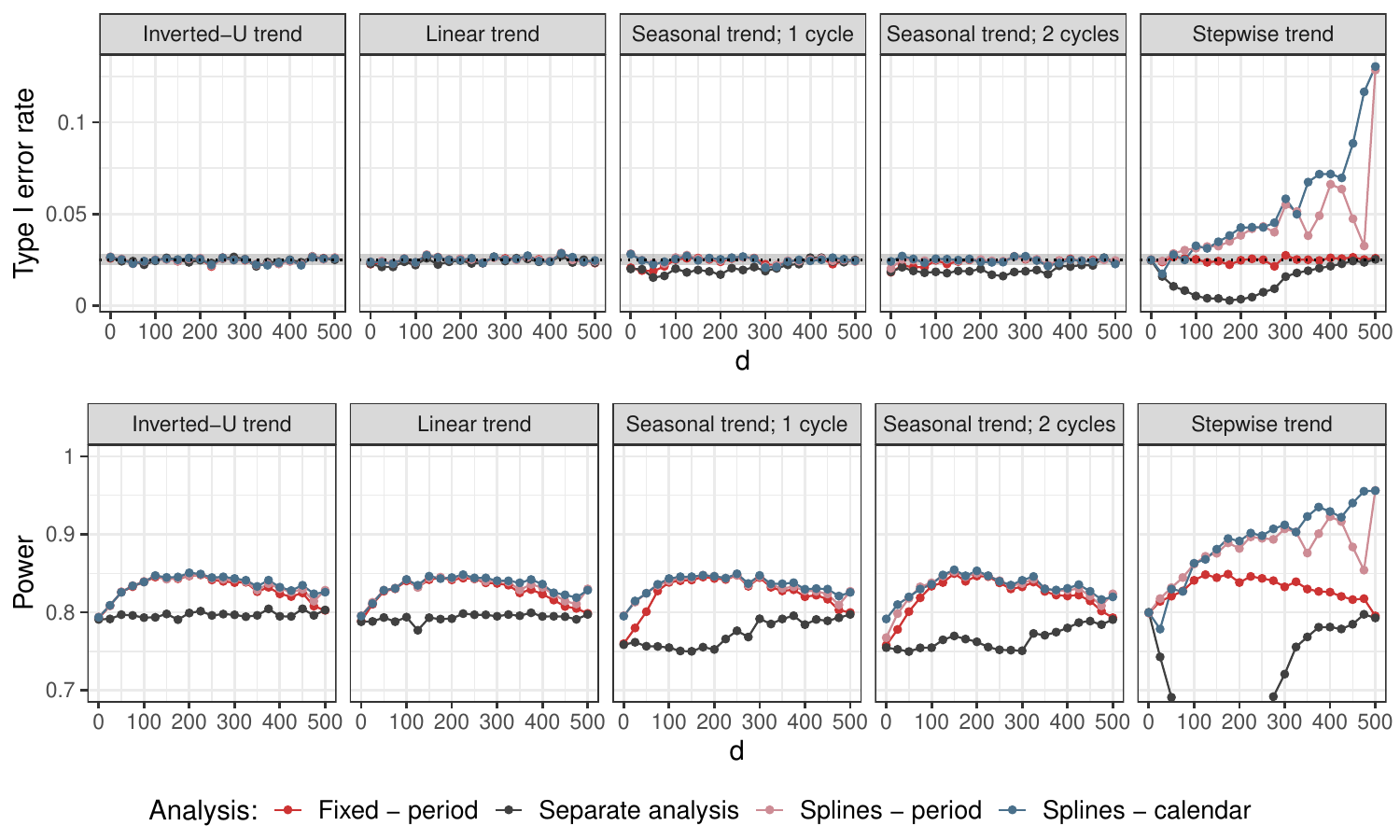}
    \caption{\textbf{Setting 1B:} Type I error rate and power for the cubic spline regression model with knots according to periods or calendar time units compared to the regression model with period adjustment with respect to the timing of adding the treatment arms using different time trend patterns. Here, $n=250$ in each experimental arm, $N_p$ placed in the middle of the trial in case of an inverted-U trend, and $\lambda=0.5$ are used. In case of calendar time adjustment, unit size of 450 patients is considered. Treatment arm 5 is being evaluated. In settings with inverted-U, linear, and seasonal time trends, the lines for splines with knots placed according to periods (pink) and calendar time units (blue) are nearly overlapping.
    }
    \label{fig:splines_d_trend}
\end{figure}

\begin{figure}[h!]
    \centering
    \includegraphics[width=1\textwidth]{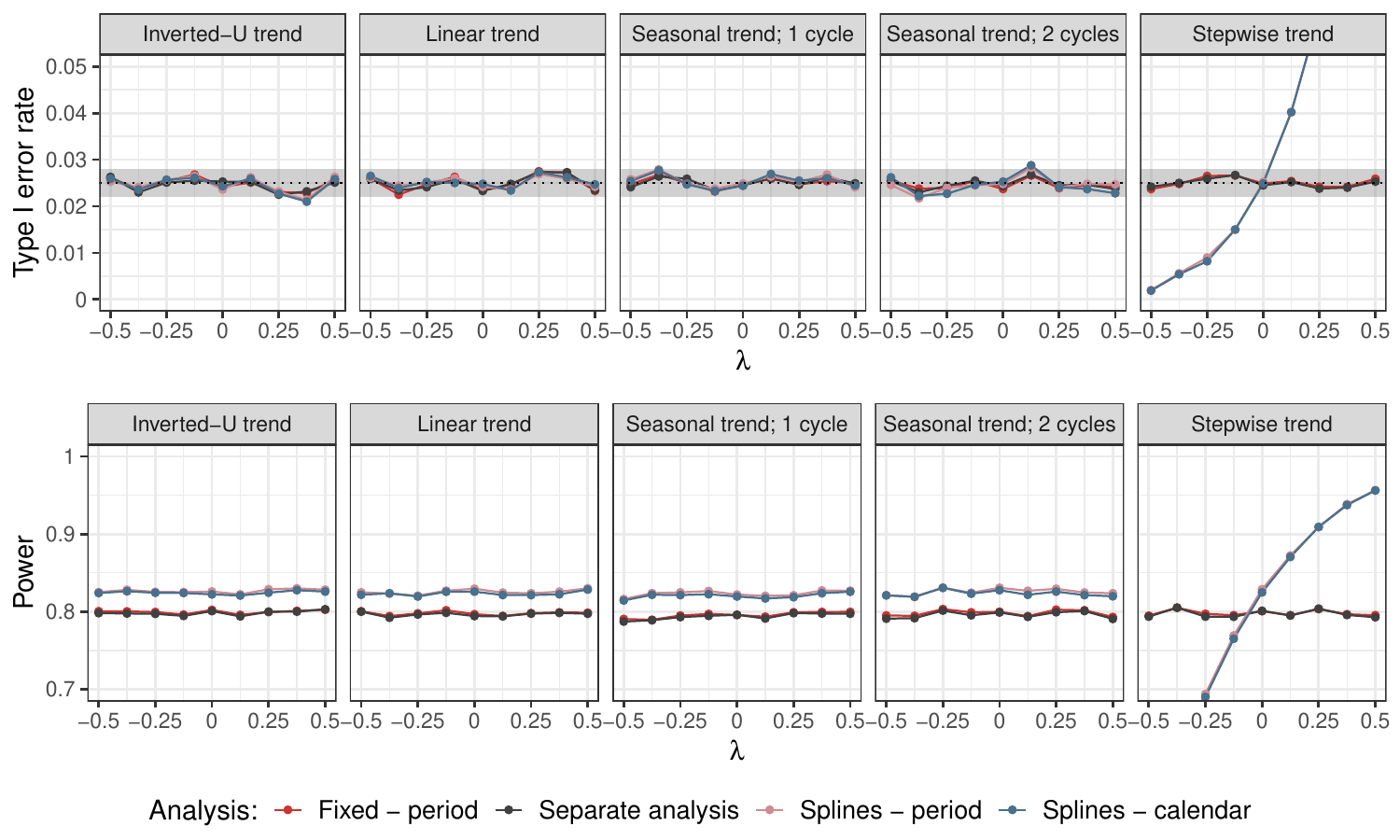}
    \caption{\textbf{Setting 1B:} Type I error rate and power for the cubic spline regression model with knots according to periods or calendar time units compared to the regression model with period adjustment with respect to the strength of the time trend $\lambda$ using different time trend patterns. Here, $n=250$ in each experimental arm, $d=500$ (no overlap between the arms), and $N_p=2500$ in case of inverted-U trend (corresponding to the middle of the trial) are used. In case of calendar time adjustment, unit size of 450 patients is considered. Treatment arm 5 is being evaluated.
    In the power plots, the lines for splines with knots placed according to periods (pink) and calendar time units (blue) are nearly overlapping, and the lines for fixed effect model with period adjustment (red) and separate analysis (black) are indistinguishable.}
    \label{fig:splines_lambda_trend}
\end{figure}

The type I error rate and power of the evaluated models under the considered scenarios when comparing treatment arm 5 to the control are presented in Figures \ref{fig:splines_d_trend} and  \ref{fig:splines_lambda_trend}. 
Analogous plots for other experimental arms are given in Figures S6 and S7 in the supplement. 
If the time trend pattern is given by a sufficiently smooth function, the spline regression maintains the type I error rate. However, in case of the stepwise time trend, we observe an inflation in the type I error rate. This is because the spline regression estimates the time effect by a smooth function, composed of multiple polynomial functions joined together in the knots. This is not a valid approach if there are sudden jumps in the time trend. The regression model maintains the type I error rate under an arbitrary time trend pattern and strength. Note that the separate approach has a type I error rate strictly below 2.5\% for strong time trends under the seasonal trend pattern and in particular under the stepwise trend pattern, as this one has stronger values than the other shapes. The reason is that block randomization was used to assign the patients to the active arms. In case of very strong time trends, ignoring the blocks in the analysis results in a conservative test, as the variances of the treatment effects are overestimated. This leads to deflation in the type I error rate, which gets stronger the more extreme the time trend is \citep{Matts1988Properties}.

In cases with less overlap between the treatment arms, the power of the spline regression is improved compared to the fixed effect model with period adjustment. In particular, in the case of no overlap ($d=500$), where the fixed effect model performs identically to the separate analysis, the spline regression achieves power gains of approximately 3 percentage points, while controlling the type I error rate for an arbitrary strength of the time trend, provided that its functional form is sufficiently smooth (see Figure \ref{fig:splines_lambda_trend}).

\clearpage

\subsubsection{Setting 2: Comparing the calendar time adjustment to the period adjustment in the fixed effect models and evaluating the mixed models without interaction}

To examine the regression model with calendar time adjustment and the mixed models without interaction, we consider a platform trial with 4 experimental treatment arms, where arm $k$ enters after $d_k = 250 \cdot (k-1)$ patients have been recruited to the trial, leading to a total sample size of 1528 patients.
We focus on evaluating treatment arm 3 and assess how the model performance depends on the pattern and strength of the time trend, as well as on the chosen calendar time length.

\begin{figure}[h!]
    \centering
    \includegraphics[width=1\textwidth]{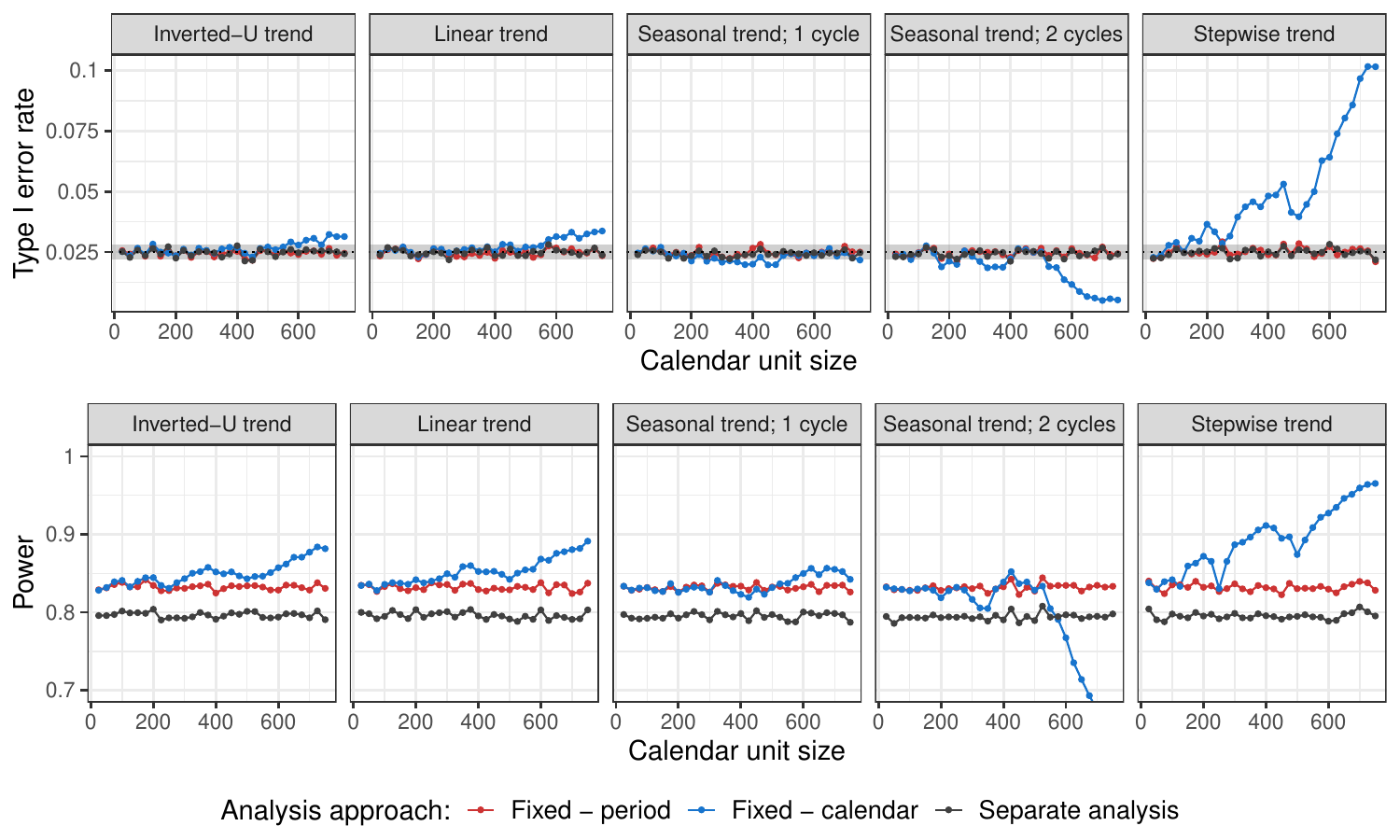}
    \caption{\textbf{Setting 2A:} Type I error rate and power of the regression model with calendar time adjustment compared to the regression model with period adjustment and separate analysis with respect to the size of the calendar time unit under different time trend patterns. Here, $n=250$ in each experimental arm, $\lambda=0.125$, and $N_p=750$ in case of inverted-U trend (corresponding approximately to the middle of the trial) are considered. Treatment arm 3 is being evaluated.}
    \label{fig:fixmodel_cal_unit}
\end{figure}

\begin{figure}[h!]
    \centering
    \includegraphics[width=1\textwidth]{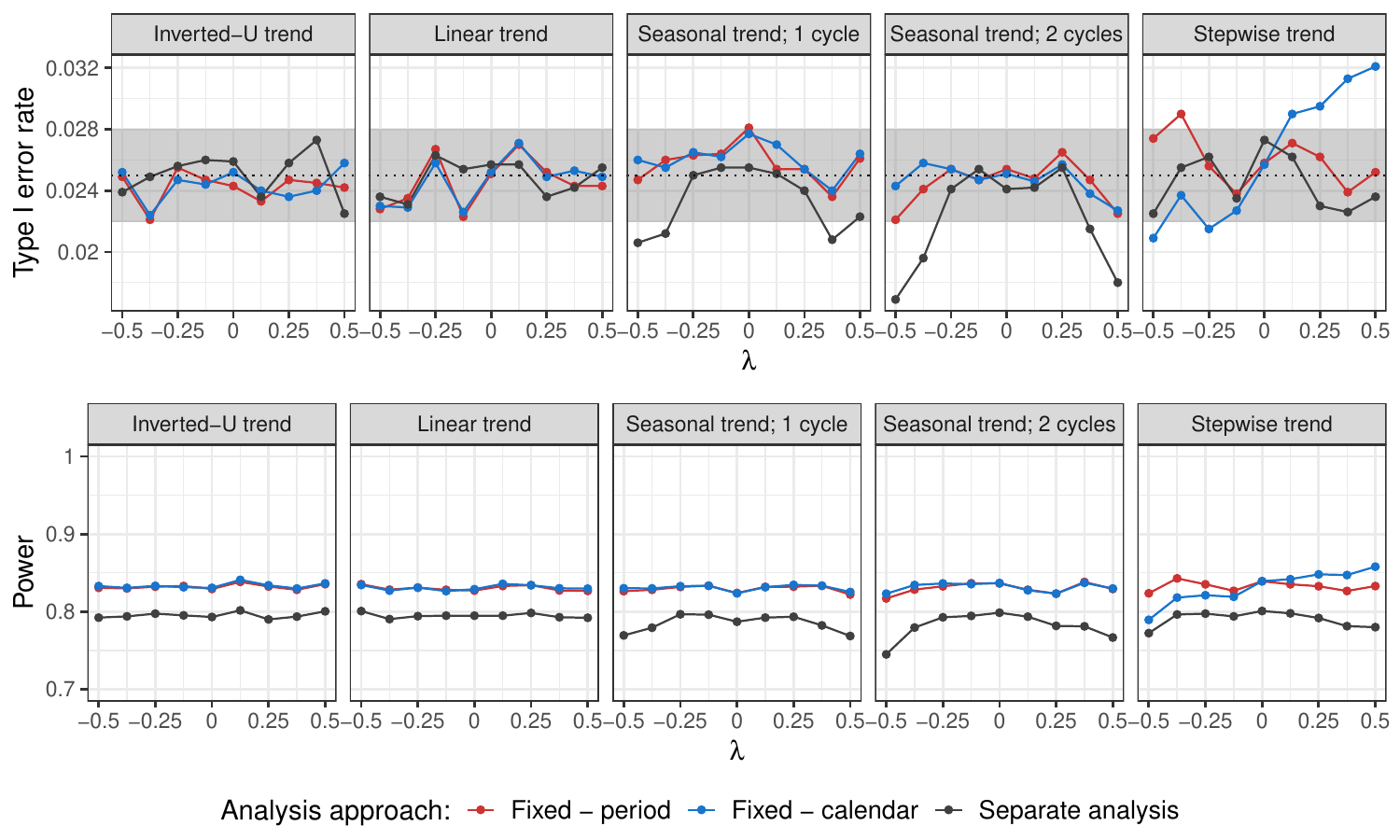}
    \caption{\textbf{Setting 2A:} Type I error rate and power of the regression model with calendar time adjustment compared to the regression model with period adjustment and separate analysis with respect to the strength of the time trend $\lambda$ under different time trend patterns. Here, $n=250$ in each experimental arm, $N_p=750$ in case of inverted-U trend (corresponding approximately to the middle of the trial), and calendar time unit size of 100 are used. Treatment arm 3 is being evaluated. In the power plots for settings with inverted-U, linear, and seasonal time trends, the lines for the regression models with period (red) and calendar time (blue) adjustment are overlapping.
    }
    \label{fig:fixmodel_cal_lambda}
\end{figure}

\textbf{Setting 2A.} In Figures \ref{fig:fixmodel_cal_unit} and \ref{fig:fixmodel_cal_lambda} the type I error rate and power of the model with the calendar time adjustment is compared to the period adjustment and the separate analysis under varying unit size and $\lambda$, respectively. 
Corresponding results for other experimental arms are presented in Figures S8 and S9 in the supplementary material. 
For Figure \ref{fig:fixmodel_cal_unit}, unit sizes in the range from 25 to 750 (see Table \ref{tab:tab_scenarios}) were considered. The length for which the type I error rate is still controlled depends on the assumed time trend pattern and its strength. In case of moderately strong linear and inverted-U trends, the type I error rate is maintained for calendar time units of approximately 1/4 of the length of the considered trial, and slightly inflated for units larger than 400 patients. Under the seasonal time trend, we observe deflation of the type I error for unit sizes larger than 200. In case of the stepwise trend, the type I error rate control is only given for very small unit sizes ($<50$ patients).

In the case considered here, with calendar time units chosen such that treatment arms can enter or leave within the interval, adjusting for these intervals leads to pooling of trial data with different randomization scheme, since the number of active arms changes. The loss of type I error rate control, which can be observed especially for larger calendar unit sizes, is then due to the greater effect of data pooling. In trials with sudden changes in the time trend (such as the considered stepwise trend pattern), the type I error inflation for naive pooling is even more pronounced.

Depending on the unit size, the model with calendar time adjustment can lead to power improvements as compared to the period adjustment. In particular, in trials where the patient response does not change substantially over time (i.e. trials with linear or inverted-U time trend pattern and small $\lambda$ relative to the treatment effect), power gains can be achieved by considering calendar time units larger than periods, while still maintaining the type I error rate control (see Figure \ref{fig:fixmodel_cal_unit}).

As the choice of the length of the calendar time interval determines the resulting trade-off between type I and type II error rates and needs to be carefully assessed in simulations when planning the platform trial. If the experimental arms would be allowed to enter and leave the trial only at the beginning of a calendar time unit, the type I error rate control would be ensured regardless of the time trend pattern and strength, as every change in the randomization design would be adjusted for. However, under this setting, the model with calendar time adjustment does not offer any gain in power as compared to the period adjustment. We did not consider this design in our simulations, as we aimed to evaluate the general case with arbitrary unit sizes.

\textbf{Setting 2B.} Similarly, for the evaluation of the mixed models without interaction, we focus on evaluating treatment arm 3, consider varying time trend patterns and strengths, and present the resulting type I error rate and power in Figure \ref{fig:mixmodel_alpha_pow}.
Figure S10 in the supplementary material shows corresponding results for other treatment arms.

We observe that the mixed models only maintain the type I error rate if no time trends are present in the trial (i.e., $\lambda=0$). Under this assumption, they also lead to power improvement compared to the fixed effect regression model. However, in case of time trends the type I error rate control is lost. The inflation (or deflation) of the type I error is most pronounced in the mixed model that adjusts for calendar time intervals as uncorrelated random effects. This is because the variance of the random effects is underestimated since fewer observations are available in each calendar time interval, and the observations within one interval are more similar to each other than it is the case for the periods. Moreover, especially in settings with seasonal and stepwise trends, the maximum inflation seems to be achieved for moderately strong time trends. The reason for this is that the mixed models shrink the effect of the time trend, since it is modeled as a random effect. If, however, the trend is large enough, there is less shrinkage and the time effect is preserved and better adjusted for. 

\begin{figure}[h!]
    \centering
    \includegraphics[width=1\textwidth]{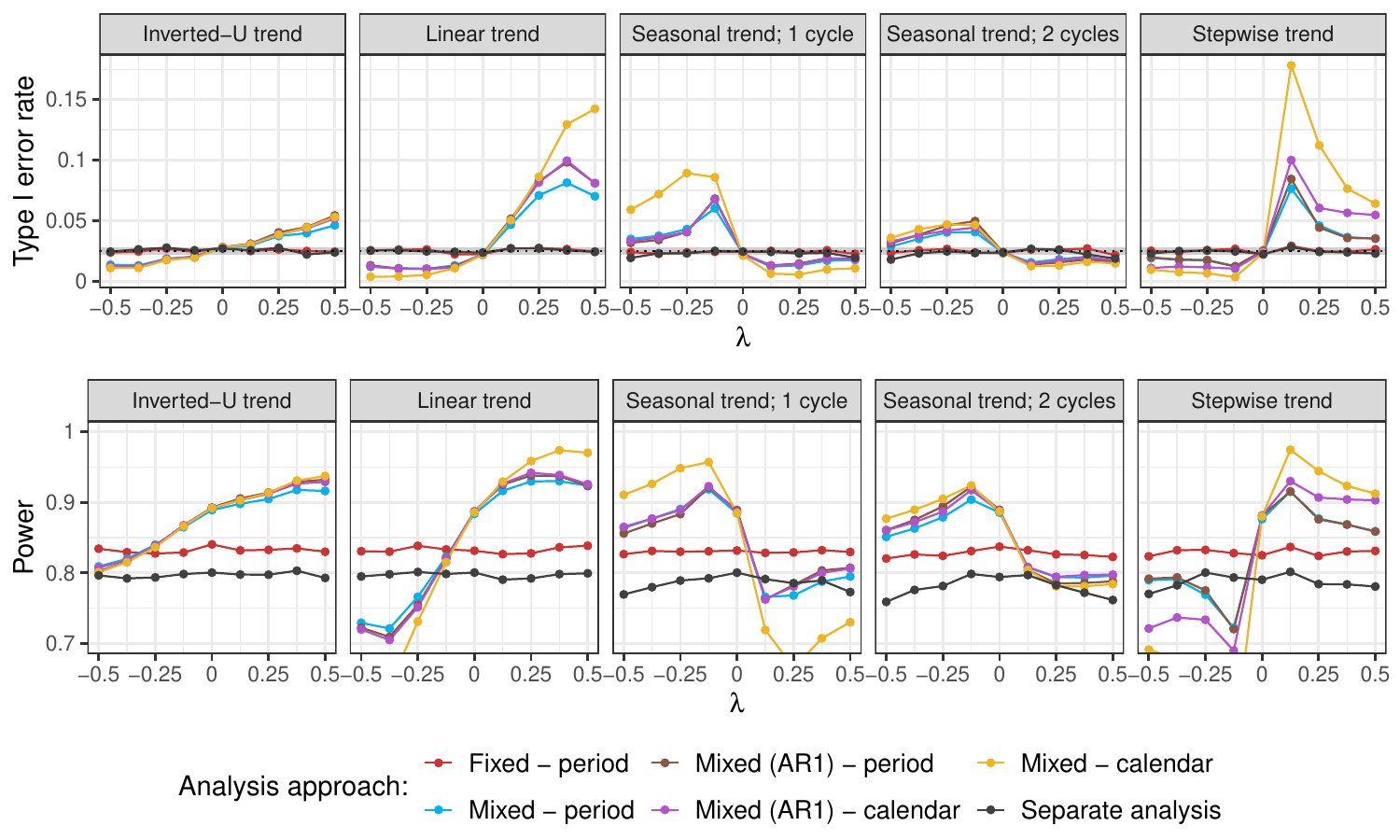}
    \caption{\textbf{Setting 2B:} Type I error rate and power of the mixed model with period and calendar time adjustments as uncorrelated and autocorrelated random effects, compared to the fixed effect regression model with period adjustment with respect to the pattern and strength of the time trend. Sample sizes of $n=250$ in each experimental arm and $N_p=750$ in case of inverted-U trend (corresponding approximately to the middle of the trial) are assumed. In case of calendar time adjustment, unit size of 100 patients is considered. Treatment arm 3 is being evaluated. In settings with inverted-U, linear, and seasonal time trends, the lines for the mixed effects models with autocorrelated random effects and period (brown) and calendar time (magenta) adjustment overlap. In the setting with stepwise time trend, the lines for the mixed effect models with period adjustment and uncorrelated (blue) and autocorrelated (brown) random effects overlap.
    }
    \label{fig:mixmodel_alpha_pow}
\end{figure}

\clearpage

\subsubsection{Setting 3: Evaluating the performance of the mixed models with interaction under different time trends}

To evaluate whether the mixed models with interaction reduce the type I error rate inflation under different time trends, we consider a scenario with $K=4$ experimental arms where treatment arm 3 is being evaluated, as in Setting 2. In this case, only the linear pattern of the time trend is considered. However, we simulate different cases with regard to the strength of the time trend. In particular, we vary the strength of the time trend in arm 1 ($\lambda_1$), arms 1 and 2 ($\lambda_1 = \lambda_2$) or arms 1, 2 and 4. In the last case, we additionally distinguish between two variants - equal time trend strengths in arms 1, 2, and 4 ($\lambda_1 = \lambda_2 = \lambda_4$) and different strengths across these arms ($\lambda_2 = 2 \lambda_1$, $\lambda_4 = 3 \lambda_1$). The remaining arms have no time trend. For comparison, we also consider a case with equal time trends in all arms ($\lambda_1 = \lambda_k, \forall k$).

\begin{figure}[h!]
    \centering
    \includegraphics[width=1\textwidth]{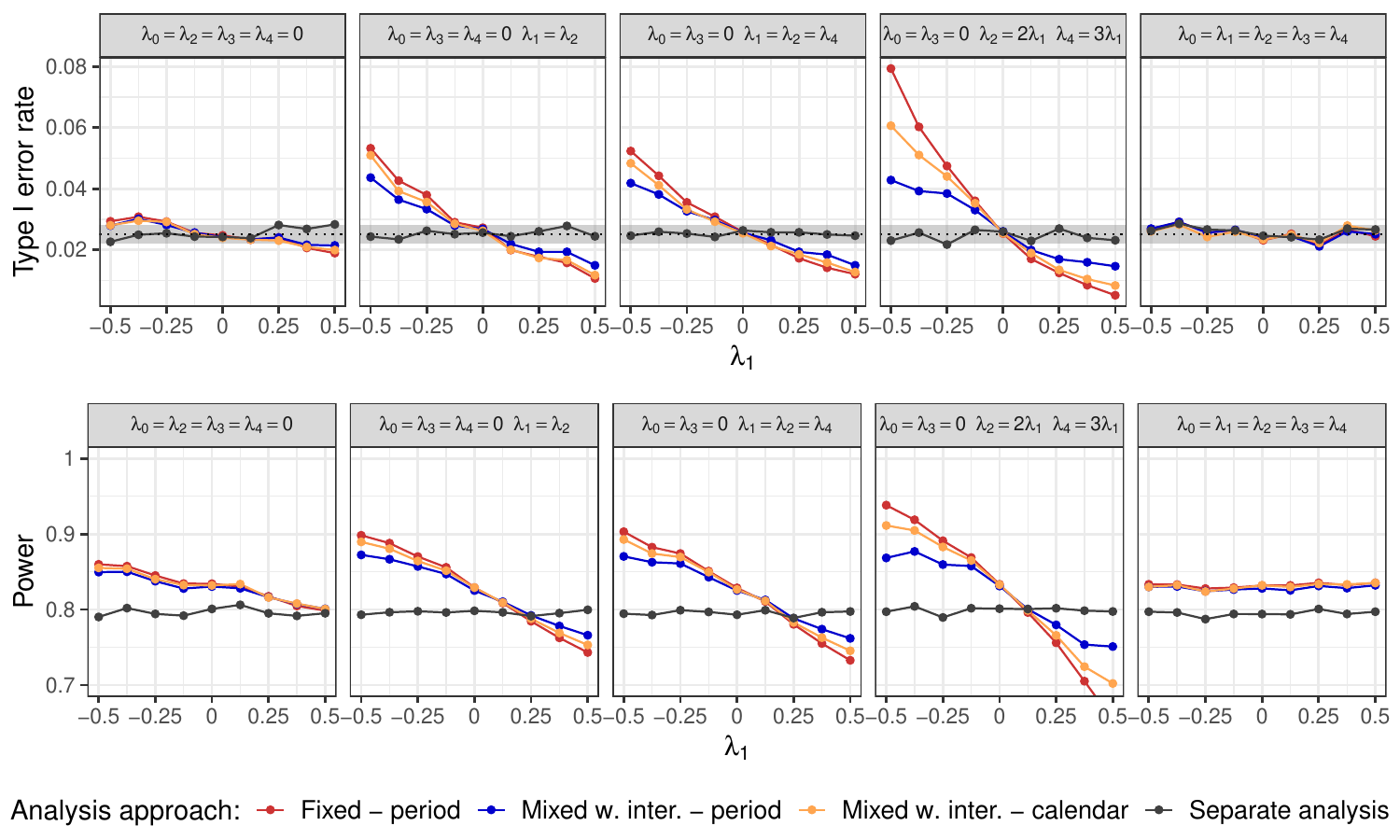}
    \caption{\textbf{Setting 3:} Type I error rate and power of the mixed model with interaction and period or calendar time adjustments, compared to the fixed effect regression model with period adjustment with respect to the time trend case and its strength. Sample sizes of $n=250$ in each experimental arm are assumed. In case of calendar time adjustment, unit size of 100 patients is considered. Treatment arm 3 is being evaluated. In the power plot under equal time trends, the lines for all model-based approaches (red, blue, and yellow) overlap.}
    \label{fig:mixint_alpha_pow}
\end{figure}

Figure \ref{fig:mixint_alpha_pow} shows the obtained type I error rate and power with respect to $\lambda_1$ for the two variants of the mixed model with interaction (adjusting for period or calendar time units), compared to the fixed effect model with period adjustment and the separate analysis. Different cases regarding the strength of the time trend are presented in the columns. The results show that the mixed models lead to less inflation than the fixed effect model in the case of different time trends. The improvement in the type I error rate inflation when using the mixed models compared to the fixed effect model is particularly pronounced in the case with more variation in the time trend strength across arms (see Figure \ref{fig:mixint_alpha_pow}, 4th column). However, as expected, the mixed effect models do not maintain the type I error rate at the nominal level for arbitrary strengths of the time trend. In the case of equal time trends, they control the type I error rate and lead to comparable power as the fixed effect model with period adjustment. 

As shown in \cite{Bofill2022Model}, the type I error rate is controlled for different time trends across arms when an interaction term is included as a fixed effect. However, such a model offers no power gains compared to the separate analysis.

\subsection{Summary of main results}

Below we summarize the advantages and limitations of the considered methods and the main results from the simulation study and describe the influence of certain design parameters.

\textbf{Fixed effect regression with period adjustment:} Fixed effect models with period adjustment maintain the type I error rate and lead to a power increase compared to the separate analysis in case of equal time trends between arms that are additive on the model scale, for arbitrary time trend patterns. However, a certain overlap in time between the experimental treatment arms is required in order to achieve a power gain. The amount of the power increase depends on the size of the NCC data and the amount of overlap.

\textbf{Spline regression with period adjustment:} Spline regression maintains the type I error rate in scenarios where the time trend pattern is sufficiently smooth and equal across all arms. Moreover, it can lead to a power increase as compared to the fixed effect model with period adjustment, especially in cases with little to no overlap between the experimental treatment arms.
    
\textbf{Mixed effect regression with period adjustment:} Mixed models where the period is modeled as random effect do not control the type I error rate in the presence of time trends. The inflation is more pronounced in cases with moderate time trends compared to cases with strong time trends. Mixed models with a fixed effect period and a random effect for the interaction between treatment and period control the type I error rate in trials with equal time trends, and reduce the inflation in settings with unequal time trends, but do not eliminate it completely. In cases with unequal time trends, the type I error rate is controlled by a regression model that includes an interaction between treatment and time as a fixed effect \citep{Bofill2022Model}. This method, however, offers no power gains compared to the separate analysis.

\textbf{Period vs. calendar time adjustment:} Additionally to the period adjustment investigated in previous works, we considered to adjust for calendar time units that are, unlike the periods, independent of the trial progress. Using this adjustment, the number of factor levels in the resulting regression model, or number of inner knots for the spline function, can be controlled by the additional design parameter $c_{length}$ (calendar time unit length). Depending on the pattern and strength of the time trend, adjusting for larger units than those given by the periods can leads to power gains, while still controlling the type I error rate. Type I error rate control would be ensured in cases where new experimental arms only enter and leave the trial at the beginning of a calendar time unit, but this setting would not lead to substantial power gains as compared to the period adjustment. Also note that the last calendar time interval is cut at the time point where the evaluated arm leaves the platform to avoid including future data to the analysis, which might not be available by the time of assessing the given arm.

\textbf{Overlap between experimental arms:} We showed that certain overlap between the experimental arms is necessary for the fixed effect regression model with period adjustment to achieve power gains as compared to the separate analysis. This is, however, not required by the spline regression, which uses continuous, rather than categorical time adjustment and leads to power increase even in cases with no overlap, i.e. if only one experimental arm is active at any given time.


\section{Case Study}\label{sect_casestudy}

We exemplify the use of the proposed methods in a case study based on a real-world clinical trial. It revisits real data from a trial for progressive supranuclear palsy \citep{Hoeglinger2021Safety} and analyzes the baseline measurements, hence illustrates the use of the modeling approaches in a setting where the null hypothesis is assumed to hold.

Additionally, in the supplementary material (Sect.~D), we present a second case study that mimics the design of the FLAIR platform trial for chronic lymphocytic leukemia \citep{Munir2024Chronic, Hillmen2023Ibrutinib, Howard2021Platform} using simulated data. It incorporates certain design elements from the FLAIR trial and demonstrates the application of methods in a scenario involving two efficacious treatment options.

We reanalyze data from the ABBV-8E12 trial \citep{Hoeglinger2021Safety} for progressive supranuclear palsy (PSP), a rare neurodegenerative disorder affecting balance, body movements, vision, and speech, eventually resulting in death. The ABBV-8E12 study evaluated the safety and efficacy of different doses (2000mg - here denoted as arm 1; and 4000mg - denoted as arm 2) of the experimental treatment Tilavonemab in a randomized parallel group trial, using placebo as a common control group. The trial recruited patients over a period of approximately 2 years, from January 2017 to March 2019. When revisiting this data, we consider the Progressive Supranuclear Palsy Rating Scale (PSPRS) as an endpoint, a sum score summing up the values for 28 different items, which measure the progression of the disease across different domains (lower values of this score indicate better outcomes) \citep{Golbe2007Clinical}. To demonstrate the potential impact of time trends under the null hypothesis, for illustrative purposes we compare the baseline measurements of the sum score between treatment and control groups (instead of the change from baseline as in the original trial). For the baseline measurements, we know that due to randomization there is no treatment effect compared to placebo for both arms, such that all observed treatment effects are either due to bias or sampling error. We furthermore demonstrate the use of the proposed models on this data. 

\begin{figure}[!h]
    \centering
    \includegraphics[width=0.8\textwidth]{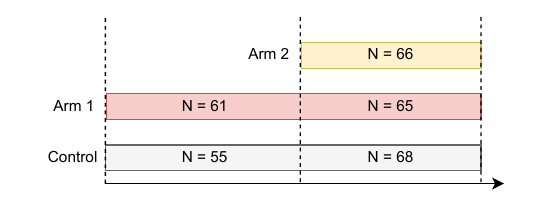}
    \caption{Scheme of the platform trial considered in the case study for progressive supranuclear palsy (ABBV-8E12 trial). Sample sizes per arm and period are shown.}
    \label{fig:case_study_scheme_Abbvie}
\end{figure}

\begin{figure}[!h]
    \centering
    \includegraphics[width=\textwidth]{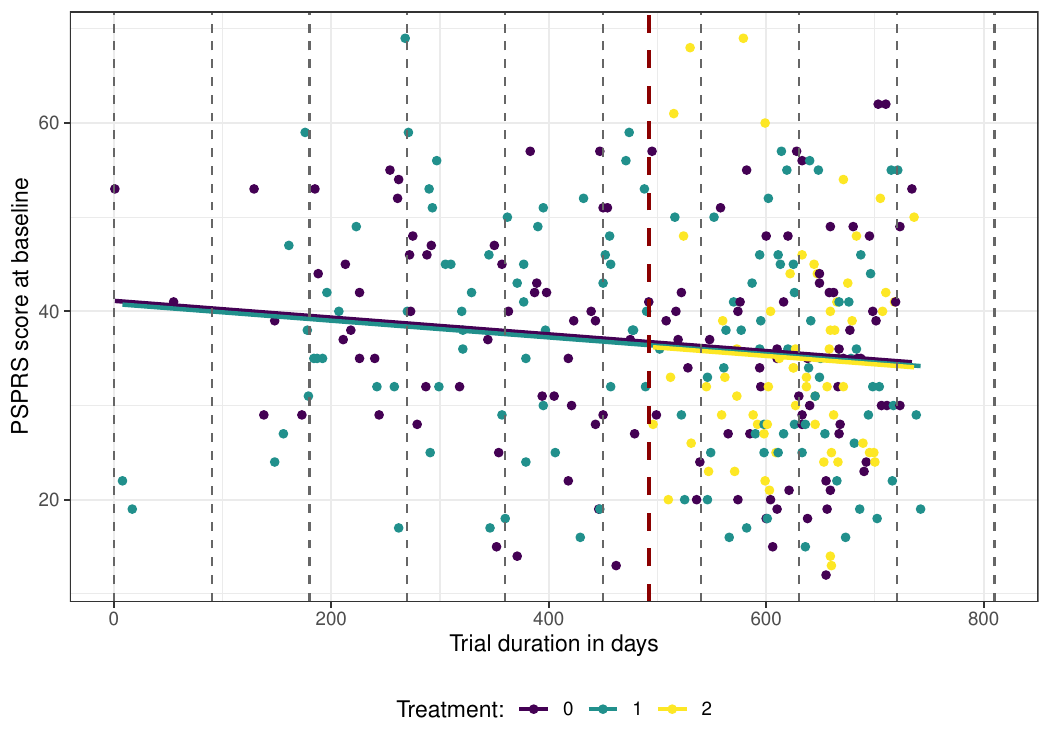}
    \caption{Data from the ABBV-8E12 trial: responses measured at baseline. The data was modified to resemble a platform trial consisting of two periods (indicated by the red dashed line) and nine calendar time units (denoted by the grey dashed lines).}
    \label{fig:case_study}
\end{figure}

To emulate a two-period platform trial, we modified the original multi-arm trial data and assumed that the second experimental arm (4000mg of Tilavonemab) enters the ongoing trial at a later time point. Therefore, we set the cut-off for the periods (i.e., the time point where treatment arm 2 enters the platform) to 1.6.2018 and exclude observations from patients allocated to arm 2 prior to this cut-off date from the analysis. The resulting dataset consists of 315 patients, with sample sizes of 123, 126, and 66 for the placebo arm, and treatment arms 1 and 2, respectively. The sample sizes per arm and period are presented in Figure \ref{fig:case_study_scheme_Abbvie}. For approaches that adjust for calendar time, we use a unit length of 90 days, which divides the trial into 9 calendar time units, each of them corresponding to approximately 3 months. Note that as the patient recruitment was not uniform, the numbers of patients in each unit are different, ranging from 4 in the first unit to 96 in the 8th. Figure \ref{fig:case_study} visualizes the considered data, showing the division into periods (red dashed line) and calendar time units (grey dashed lines). 

First, we investigate the presence of a possible time trend in the modified trial data by fitting a linear model to the primary outcome, using treatment as categorical, and days since the recruitment of the first patient as continuous covariates. Fitted values from this model are included as solid lines in Figure \ref{fig:case_study}. The model showed a significant negative time trend (p=0.032), which is also visible in the figure. Hence, at the beginning, the trial recruited patients with more severe stage of the disease, while participants recruited later had on average a milder stage of PSP. 

Subsequently, we analyze the dataset using the proposed modeling approaches, as well as the separate and pooled analyses, and compare the estimated treatment effects, their standard errors, and p-values. In the case of the mixed models, we only consider the calendar time adjustment, as the analyzed trial only consists of two periods, which would lead to a very unstable estimation of the random effect in case of the period adjustment. We focus on evaluating the efficacy of the 2nd experimental arm against control, since this arm has non-concurrent control data available. Table \ref{tab:tab_case_study} summarizes the obtained results. Since the considered responses were measured at the baseline date in a randomized trial, we can assume that the null hypothesis holds, thus the underlying treatment effect is 0 for both arms. We evaluate this null hypothesis ($H_{0,2}: \theta_2 = 0$) against the two-sided alternative ($H_{1,2}: \theta_2 \neq 0$). None of the analysis approaches reject the null hypothesis. The most accurate estimate of the treatment effect (-0.012) is provided by the fixed effect model adjusting for periods, followed by the fixed effect model with calendar time adjustment (-0.068). These models also result in the highest p-values against the null hypothesis (p=0.995 and p=0.971, respectively). The second closest estimate is provided by the spline regression, resulting in estimated effects of 0.178 and 0.624 for the period and calendar time adjustment, respectively. In both model classes (fixed effect models and spline regressions), the period adjustment leads to a better effect estimate than the calendar time adjustment. The standard errors of the estimates are comparable for all considered models, ranging from 1.822 (mixed model with autocorrelated random effects and calendar time adjustment) to 1.900 (fixed effect model with calendar time adjustment). We can also observe that all modeling approaches yield estimates that are closer to the true effect than the pooled analysis, which introduces a considerable bias to the effect estimation (-1.7030) and also provides the lowest p-value (p=0.336). 

In the ABBV-8E12 trial, the primary endpoint was the change from baseline in the PSPRS score after 52 weeks. We also investigated potential time trends for this endpoint, but observed no substantial drift in this outcome during the trial (data not shown). Hence, this data suggests that time trends would not pose a problem in this particular trial, and the fact that patients with a milder stage of PSP were recruited later in the trial is accounted for by adjusting for the baseline response. For an assessment of the efficiency of other choices for the primary endpoint, see \cite{yousefi2023efficiency}.

\begin{table}[]
\centering
\begin{tabular}{|c|c|c|c|c|}
\hline
\textbf{Analysis approach} & \textbf{Adjustment} & \begin{tabular}[c]{@{}c@{}}\textbf{Effect estimate}\end{tabular} & \textbf{Std. error} & \textbf{p-value} \\ \hline
\multirow{2}{*}{Fixed effect model} & Periods & -0.012 & 1.896 & 0.995 \\ \cline{2-5} 
 & Calendar time units & -0.068 & 1.900 & 0.971 \\ \hline
\multirow{1}{*}{Mixed model} & Calendar time units & -0.903 & 1.833 & 0.623 \\ \hline
 \multirow{1}{*}{Mixed model (AR1)} & Calendar time units & -0.973 & 1.822 & 0.594 \\ \hline
\multirow{2}{*}{Spline regression} & Periods & 0.178 & 1.885 & 0.924 \\ \cline{2-5} 
 & Calendar time units & 0.624 & 1.879 & 0.740 \\ \hline
Pooled analysis & - & -1.703 & 1.766 & 0.336 \\ \hline
Separate analysis & - & -0.410 & 2.032 & 0.840 \\ \hline
\end{tabular}
\caption{ABBV-8E12 trial: Treatment effect estimates, standard errors and two-sided p-values obtained when comparing the second treatment arm against the control using the proposed approaches.}
\label{tab:tab_case_study}
\end{table}

\clearpage

\section{Discussion}\label{sect_discussion}

The use of non-concurrent controls in the analysis of platform trials has been the subject of discussions and intense methodological research during recent years.
In this work, we extended the currently available frequentist methods and proposed novel, model-based approaches to utilize non-concurrent controls for individual treatment-control comparisons.
We introduced an alternative definition of the time covariate in the frequentist models, where the duration of the trial is divided into calendar time intervals of equal length. This is analogous to the time buckets considered in the Bayesian ``Time Machine'' approach \citep{Saville2022Bayesian}. Moreover, we considered more flexible means of modeling the time trend. In particular, we proposed adding the time covariate as a random effect in a linear mixed model. In these models, we additionally allowed for autocorrelation between the random effects in order to account for dependency between closer time intervals. Furthermore, we employed B-splines to model time with a polynomial function in a spline regression. Here we considered two options of placing the inner knots - either according to the periods in the trial or according to the calendar time intervals.
To relax the assumption of equal time trends crucial for all previously considered methods, we proposed an extension of the modeling approaches, where we included treatment and time as fixed effects and the interaction between them as a random effect.

We evaluated the performance of the proposed methods in terms of the type I error rate and statistical power in a simulation study, considering a wide range of scenarios. We specified under which conditions the fixed effect model adjusting for periods leads to power gains as compared to the separate analysis. Moreover, we showed that the calendar time adjustment in fixed effect models can, in some scenarios, lead to power gains as compared to the period adjustment while still controlling the type I error rate. In the considered scenarios, the mixed models without interaction were not suitable for evaluating the efficacy of late-entering treatment arms in the presence of time trends, as they do not preserve the type I error rate. \cite{Lee2020Including} and \cite{Bofill2022Model} investigated a linear regression model that includes time as a continuous, rather than categorical variable. It was shown that this model only controls the type I error rate if the underlying time trend has a linear pattern. The spline regression can be viewed as a generalization of this model, fitting one polynomial function for each period or calendar time interval, rather than one linear function for the whole trial, to adjust for time. We demonstrated via simulations that modeling time via a spline function controls the type I error rate for all considered scenarios with sufficiently smooth time trends. Another method that uses smoothing across time to estimate the treatment effect is the Bayesian Time Machine \citep{Saville2022Bayesian}. However, the performance of the Time Machine strongly depends on the choice of prior distributions. Moreover, it requires the specification of the calendar time intervals. The spline regression with period adjustment, where the placement of the knots is given by the periods, on the other hand, only requires the choice of the degree of the polynomial function, with cubic splines being a suitable default option, as they ensure type I error rate control even in cases with complex time trend patterns. Furthermore, it was shown that the models with fixed factor period that also include a term for the interaction between treatment and time as a random effect, can improve the robustness and decrease the inflation of the type I error rate compared to the model without this interaction, if the time trends are not equal across arms. The proposed analysis approaches focus only on trials with continuous endpoints. The extension of the presented methods to other types of endpoints, such as binary or time-to-event endpoints, remains open for future research. 

Non-concurrent controls share some characteristics with historical controls, as they both refer to data collected prior to the data on the treatment under study, and thus both might introduce bias in the estimates and inflate the type I error rate if included in the analysis \citep{Burger2021Use, Kopp2020Power}. However, compared to historical controls, non-concurrent controls comprise patients who have been part of the same trial framework as the patients allocated to the investigated treatment. Therefore, they typically share the inclusion and exclusion criteria and endpoint assessment procedures \citep{Sridhara2022Use}. 
Methods for incorporating historical controls in randomized clinical trials have been widely discussed in the last years, and could be analogously used in the context of platform trials to incorporate NCC data \citep{Bofill2022Review}. Frequentist methods include for instance the ``test-then-pool'' approach, where either separate or pooled analysis is performed based on a frequentist test assessing the equality of the CC and historical data \citep{Viele2014Use}; dynamic pooling, which assigns a weight parameter to the historical data to control their proportion in the analysis \citep{Jiao2019Utilizing}; or propensity score methods that can be used to adjust for imbalances in baseline covariates between historical and concurrent controls \citep{Schmidli2020Beyond}. Bayesian approaches encompass the power prior method, which down-weights the historical information by introducing a power parameter to their likelihood function \citep{Banbeta2019Modified}; commensurate power prior, an extension to the power prior approach that directly parameterizes the commensurability of the historical and CC data \citep{Hobbs2011Hierarchical}; and meta-analytic-predictive (MAP) prior approaches, which uses the historical information to derive a MAP prior distribution for the CC data \citep{Schmidli2014Robust, Weber2021Applying}. Other Bayesian methods, such as dynamic borrowing, were also proposed for incorporating external data into the trial analysis in order to obtain more information for the decision-making \citep{Edwards2024Using, DiStefano2023Incorporation}. However, in contrast to the modeling approaches to incorporate NCC data, these methods do not directly adjust for the time trend, but rather down-weight the historical information in order to ameliorate the bias due to heterogeneity in the data. Hence, they still introduce some bias if time trends are present. On the contrary, the model-based approaches for incorporating NCC lead to unbiased estimates provided that the model assumptions are met. A detailed overview of methods discussed in the context of historical, non-concurrent, and/or external controls is provided in the scoping review by \cite{Bofill2022Review}.

Recently, \cite{marschner2024analysis} considered the use of NCCs in trials with varying eligibility criteria across arms. In this case, also the use of concurrent control subjects that did not meet the eligibility criteria of the corresponding treatment arm may introduce bias in treatment control comparisons.  
They refer to non-concurrent controls for a given experimental arm as patients who do not belong to the same concurrently randomized cohort. The concurrently randomized cohort comprises only patients that were eligible for the same set of treatments and were randomized using the same allocation method \citep{marschner2024analysis}.
In this paper we assumed that for all arms the same eligibility criteria apply. If this is not the case, one can fit the proposed models in the subgroup of patients that satisfy the eligibility criteria of the tested treatment arm. 

In addition to the risk of bias due to time trends, there might be other sources of bias when utilizing non-concurrent controls. As platform trials are designed to run over multiple years, results from completed arms could be published before the whole trial ends. This may influence the specification of the analysis methods and thus the trial integrity for arms that are still active or are being considered to enter the platform \citep{Koenig2024Current}. For instance, it may affect the willingness of the intervention owners to include new treatment arms in the platform, making platform trials with a control with a random low in the outcome more attractive to join. Conversely, a platform trial with a random high can be a deterrent for the intervention owners.

While analysis using concurrent controls only is still considered the preferred approach, under certain circumstances, such as trials for rare diseases, the use of non-concurrent controls can be valuable \citep{FDA2023Master, Bofill2022Review}. In such cases, methods for addressing the potential bias due to time trends should be employed \citep{FDA2023Master}. The proposed models in this work aim to address these time trends and can improve the precision of the treatment effect estimators, while mitigating the biases that could arise due to temporal changes. However, given that the performance of the analysis methods depends on many factors, the implications of including non-concurrent controls and the behavior of the considered methods must be investigated on a case-by-case basis when planning a specific platform trial.

\vspace*{2pc}
\noindent {\bf{Supplementary Material}}
Formal definition of the time variables, further specifications of the models, and additional results from the simulation study can be found in the pdf file with supplementary material. 

The GitHub repository (\url{https://github.com/pavlakrotka/NCC_FreqModels}) contains the R code to reproduce the results presented in this paper.
	
\vspace*{1pc}
\noindent {\bf{Authors' Contributions}}
M.B.R., M.P. and P.K. conceived this research. G.H. and M.G. established data access and harmonized data across datasets. P.K. prepared the initial draft of the text and performed the simulations and statistical analyses. All authors discussed the results, provided comments and reviewed the manuscript.

\vspace*{1pc}

\noindent {\bf{Acknowledgments}}
The authors appreciate the insightful comments and suggestions on the application in Section 5 by the IMPROVE-PSP consortium members Mats O. Karlsson and Franz König. This research was funded in whole, or in part, by the Austrian Science Fund (FWF) [ESP 442 ESPRIT-Programm]. For the purpose of open access, the author has applied a CC BY public copyright licence to any Author Accepted Manuscript version arising from this submission. P.K. and M.P. received funding from the European Joint Programme on Rare Diseases (Improve-PSP). G.H. was funded by the Deutsche Forschungsgemeinschaft (DFG, German Research Foundation) under Germany's Excellence Strategy within the framework of the Munich Cluster for Systems Neurology (EXC 2145 SyNergy – ID 390857198) and the European Joint Programme on Rare Diseases (Improve-PSP). M.G. was funded by the European Joint Programme on Rare Diseases (Improve-PSP) and the Swedish Research Council Grant 2018-03317.

This publication is based on research using data from data contributors AbbVie that has been made available through Vivli, Inc. Vivli has not contributed to or approved, and is not in any way responsible for, the contents of this publication.


\vspace*{1pc}

\noindent {\bf{Conflict of Interest Statement}}
G.H. has ongoing research collaborations with Roche, UCB, Abbvie; serves as a consultant for Abbvie, Alzprotect, Amylyx, Aprinoia, Asceneuron, Bayer, Bial, Biogen, Biohaven, Epidarex, Ferrer, Kyowa Kirin, Lundbeck, Novartis, Retrotope, Roche, Sanofi, Servier, Takeda, Teva, UCB; received honoraria for scientific presentations from Abbvie, Bayer, Bial, Biogen, Bristol Myers Squibb, Esteve, Kyowa Kirin, Pfizer, Roche, Teva, UCB, Zambon; received publication royalties from Academic Press, Kohlhammer, and Thieme.

Other authors have declared no conflict of interest.

\vspace*{1pc}
\noindent {\bf{Data Availability Statement}}
The data from the ABBV-8E12 trial are available from Vivli - Center for Global Clinical Research Data at \url{https://vivli.org/}. Restrictions apply to the availability of these data, which were used under license for the current study, and so are not publicly available. The data from the simulation study and the code to generate the simulated data  are available in the GitHub repository \url{https://github.com/pavlakrotka/NCC_FreqModels}.

\vspace*{1pc}
\noindent {\bf{Patient Consent Statement}}
For the ABBV-8E12 trial \citep{Hoeglinger2021Safety}, whose data are used in this reanalysis for the case study, all participants and their respective study partners were required to provide written informed consent before screening or any study specific procedures. 

\vspace*{1pc}
\noindent {\bf{Ethics Approval Statement}}
The ABBV-8E12 trial received independent ethics committee or institutional review board approval at each study site before initiation. The study adhered to all applicable local regulations, was done in accordance with Good Clinical Practice, as outlined by the International Conference on Harmonisation, and complied with ethical standards described in the Declaration of Helsinki. See \cite{Hoeglinger2021Safety} for further details. The study protocol for the reanalysis presented in this paper was approved by the ethics committee of the Hannover Medical School, Hanover, Germany (No 9400 BO K 2020) and all methods were performed in accordance with the relevant guidelines and regulations.

\vspace*{1pc}
\noindent {\bf{Clinical Trial Registration}}
This trial is registered with ClinicalTrials.gov, NCT02985879.


\bibliography{refs}

\newpage


\includepdf[pages=-]{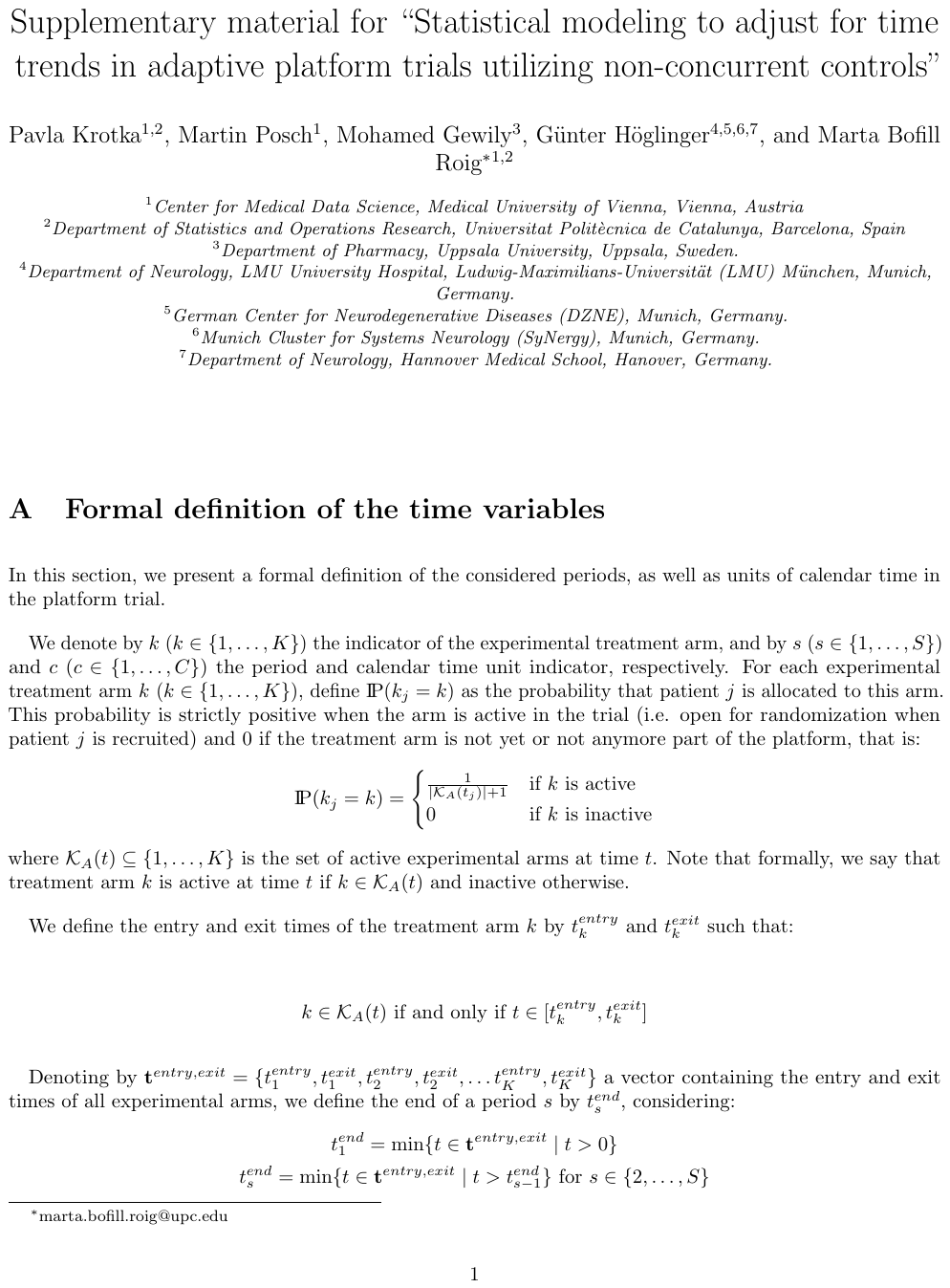}

\clearpage

\end{document}